\RequirePackage{amsmath}
\documentclass[runningheads]{llncs}

\usepackage[T1]{fontenc}

\usepackage{graphicx}
\usepackage{booktabs}
\usepackage{subcaption}
\usepackage{tabularx}
\usepackage{array}
\usepackage{numprint}
\usepackage{paralist}
\usepackage{amsmath,amsfonts,amssymb}

\usepackage{microtype}

\usepackage{xcolor}
\usepackage{soul}            % for \hl and \st
\sethlcolor{green}
\usepackage{todonotes}       % for \todo

\usepackage[hidelinks]{hyperref}
\usepackage{cleveref}

\usepackage{orcidlink}

\npthousandsep{\,}

\newcommand{\autocite}[1]{\cite{#1}}
\newcommand{\textcite}[1]{\cite{#1}}

\makeatletter
\newcommand*{\inlineequation}[2][]{%
  \refstepcounter{equation}%            % increment the equation counter
  \if\relax\detokenize{#1}\relax
  \else
    \label{#1}%                          % optional label
  \fi
  \(\displaystyle #2\)~(\theequation)%   % the actual inline math + number
}
\makeatother

\begin{document}

\title{Backdoor Attacks on Transformers for Tabular Data: An Empirical Study}
\titlerunning{Backdoor Attacks on Transformers for Tabular Data}

\author{%
  Bart Pleiter\inst{1}%
  \and
  Behrad Tajalli\orcidlink{0009-0001-0636-3385}\inst{1}%
  \and
  Stefanos Koffas\orcidlink{0000-0001-6543-4801}\inst{2}%
  \and
  Gorka Abad\orcidlink{0000-0002-6735-3623}\inst{1,3,5}%
  \and
  Jing Xu\orcidlink{0000-0002-9900-4081}\inst{2}%
  \and
  Martha Larson\orcidlink{0000-0002-1430-4721}\inst{1}%
  \and
  Stjepan Picek\orcidlink{0000-0001-7509-4337}\inst{1,4,5}%
}
\authorrunning{Pleiter et al.}

\institute{%
  Radboud University, Nijmegen, The Netherlands
  \and
  Delft University of Technology, Delft, The Netherlands
  \and
  Ikerlan Technology Research Centre, Arrasate-Mondragón, Spain
  \and 
  University of Zagreb Faculty of Electrical Engineering and Computing, Unska 3, 10000, Zagreb, Croatia
  \and
  University of Bergen, Bergen, Norway
}

\maketitle

\begin{abstract}
Deep Neural Networks (DNNs) have shown great promise in various domains. However, vulnerabilities associated with DNN training, such as backdoor attacks, are a significant concern. These attacks involve the subtle insertion of triggers during model training, allowing for manipulated predictions.
More recently, DNNs used with tabular data have gained increasing attention due to the rise of transformer models.
Our research presents a comprehensive analysis of backdoor attacks on tabular data using DNNs, mainly focusing on transformers. 
We propose a novel approach for trigger construction: \emph{in-bounds} attack, which provides excellent attack performance while maintaining stealthiness. 
Through systematic experimentation across benchmark datasets, we uncover that transformer-based DNNs for tabular data are highly susceptible to backdoor attacks, even with minimal feature value alterations. We also verify that these attacks can be generalized to other models, like XGBoost and DeepFM. 
Our results demonstrate up to $100\%$ attack success rate with negligible clean accuracy drop.
Furthermore, we evaluate several defenses against these attacks, identifying Spectral Signatures as the most effective. 
Still, our findings highlight the need to develop tabular data-specific countermeasures to defend against backdoor attacks.
\end{abstract}


\section{Introduction}
\label{introduction}

With ever more available data and processing power, Deep Neural Networks (DNNs) have firmly established their dominance in handling tasks across image, text, and audio domains, mainly consisting of homogeneous data. However, classical Machine Learning (ML) solutions like gradient-boosted decision trees are still more prevalent for heterogeneous data like tabular data~\cite{shwartzziv2021tabular}. For example, in the financial sector, decision trees are often preferred over DNNs because of their high interpretability, which is essential for regulatory compliance~\cite{shwartzziv2021tabular}. Recent studies have tried developing DNNs specifically for tabular data to improve their performance~\cite{borisov2022deep}. One of the main recent approaches outperforming the rest is transformer-based models~\cite{gorishniy2021revisiting}.

Due to the extensive data and computational resources DNNs demand for training, many users outsource the training process to third parties, employ pre-trained models, or use data sources of suspicious trustworthiness. This can lead to potential security threats, like backdoor attacks~\cite {li2022backdoor}.
In a backdoor attack, the adversary injects a hidden functionality in a model during training (usually by poisoning the dataset or model directly). During inference, the backdoored model functions normally on clean data but outputs the desired target label when facing malicious inputs that contain the backdoor trigger~\cite{li2022backdoor}.
Tabular data is highly used in critical sectors such as the financial sector or transportation, making it appealing for attackers to target. As such, backdoor attacks on tabular data can pose a realistic threat in many scenarios, e.g., in the financial sector, where an adversary intending to secure a loan manipulates the model to predict their capability to repay the loan. The adversary can influence the prediction result by slightly modifying one or a few features. 
%In other fields, such as transportation, the attacker could arbitrarily alter the schedule or routes, causing severe alterations in logistics and additional costs.

Despite extensive research on backdoor attacks, only a few have considered studying backdoor attacks on DNNs for tabular data~\cite{DBLP:conf/iclr/XieHCL20,joe2022exploiting}. 
%Nevertheless, important research questions still need to be addressed. \todo{g: which are the questions? I dont like the phrase but i dont know what to write yet}
Consequently, there are still open questions to address. We investigate how vulnerable DNNs for tabular data are against backdoor attacks. Given the unique characteristics of tabular data (see~\autoref{subsec:tabularCharacteristics}), we investigate how we can embed a hidden trigger in tabular data compared to image and text data. Moreover, we explore the most relevant parameters for a backdoor attack on tabular data and how different trigger-generation methods affect the backdoor's performance. Finally, we evaluate how backdoors on tabular data can be prevented by adapting state-of-the-art defenses from other domains, such as computer vision.

% This paper experimentally analyzes backdoor attacks and defenses on tabular data. We focus primarily on transformer-based neural networks because of their superior performance on tabular data compared to other DNNs~\cite{gorishniy2021revisiting}. However, we also verify that our attack can be applied to classical ML models and other types of DNNs.
% Due to their particular features, backdooring tabular data may require \added{unique} techniques, different from images or text. Tabular data are usually heterogeneous, and each feature can be of a different type with different statistics and distributions, thus making it impossible to apply the same backdoor trigger value for all of them. Additionally, unlike images, shuffling the order of the features in tabular data will not affect model performance, making it infeasible for backdoors to use spatial dependencies between features. \todo{g: for me this paragraph is disconnected and does not belong to introduction. I think it should be in the rationales of the attack. stef: I agree}

Our investigation covers different aspects of the security of tabular data. (i) We first evaluate the importance of different aspects of DNNs, considering which are the most successful for injecting a backdoor. (ii) We evaluate different types of attacks that vary in stealthiness: out-of-bounds, in-bounds, and clean label attacks, which do not change the samples' labels, compared to dirty label attacks.
%{We inject backdoors for tabular data by investigating which factor plays the most important role in successfully backdooring the classifier models. We also make the backdoor stealthy by using values from the dataset's distributions (i.e., in-bounds triggers) and performing clean label attacks. Clean label attacks, unlike dirty label attacks, do not change the label of the poisoned sample.}
We assess each attack using three state-of-the-art transformers on four benchmark datasets for tabular data. Additionally, we verify that these attacks can be applied to classical ML models like XGBoost~\cite{chen2016xgboost} and other types of DNNs like DeepFM~\cite{guo2017deepfm}. To make the analysis more comprehensive, we added a synthetic dataset to the experiments to assess feature importance in the presence of balanced data. Our results show that models trained on tabular data could be backdoored in almost all cases, with an attack success rate close to 100\%. 
We also adapt three defenses to detect or repair the poisoned model, of which we find Spectral Signatures most successful. 
Our main contributions are:
\begin{itemize}
    \item To our knowledge, this is the first study that comprehensively analyzes backdoor attacks on tabular data using DNN-based models.\footnote{Our code is available at~\url{https://github.com/HamidRezaTajalli/Tabdoor.git}}
    \item We are the first to consider transformer-based DNNs for tabular data in the presence of backdoors, finding them highly vulnerable to backdoor attacks. More precisely, by changing a single feature value, we achieve a high (\(\approx 100\%\)) attack success rate (ASR) with low poisoning rates on all models and datasets.
    % \item We design a novel backdoor attack for tabular data that can be applied to all tabular datasets, including numerical and categorical features, and used for classification.
    %By adapting a backdoor attack designed for Federated Learning (FL) to a centralized setup\todo{stef: Maybe this sentence downplays our work}, we create the first backdoor that can be applied to all tabular datasets that include numerical and categorical features and used for classification.
    \item We study multiple attack variations. We perform a clean label attack that can reach more than 90\% ASR in most of our experiments without having to implement any trigger optimization technique. We also propose a new attack with in-bounds trigger values that reach an ASR close to perfect (\(\approx 100\%\)) even with a very low poisoning rate.
    % \item After the poisoning rate, we find the trigger location to be the most important parameter for the backdoor. While altering features with high feature importance scores generally leads to high attack success rates, we also notice that the chosen feature, i.e., the trigger location, is not the only factor influencing the attack performance.
    \item We explore several defenses against backdoor attacks. We conjecture that detection techniques using a latent representation can be the best option for defending against these attacks. Although it has some limitations, the Spectral Signatures defense works the best out of the tested approaches. However, our adaptive attacker could bypass this defense in most cases.
\end{itemize}

\section{Background and Related Work}
In deep neural networks (DNNs), a backdoor embeds a hidden functionality during training so that malicious inputs force the model to predict an attacker-chosen target label $y_t$ while maintaining normal performance on clean data. These triggers are most commonly introduced via data poisoning—injecting a small fraction of manipulated samples into the training set—though code alteration and weight poisoning are also possible.  Attackers aim for stealth by keeping the poisoning rate low and ensuring minimal effect on clean accuracy, causing the model to learn trigger-specific feature activations that yield high-confidence target predictions only when the trigger is present.

Recently, backdoors in tabular data were introduced, where heterogeneous feature types and complex interactions present unique challenges. Classical methods like gradient-boosted decision trees often outperform vanilla DNNs, but transformer-based architectures (e.g., TabTransformer, SAINT) use self-attention to capture feature relationships more effectively. Prior studies have demonstrated the feasibility of backdoors on electronic health records via covariance- or autoencoder-based triggers~\cite{joe2022exploiting}, and on federated learning, by distributing feature-based triggers across participants~\cite{DBLP:conf/iclr/XieHCL20}. However, these efforts focus on time-series or distributed settings and do not address vulnerabilities in transformer-based tabular models, motivating our systematic analysis in this domain.

\section{Attack Rationale \& Setup}
\label{sec:attack_setup}

In this section, we outline the framework for conducting backdoor attacks on tabular data, emphasizing the unique attributes posed by its heterogeneous nature. We formally define the attacker's objectives, knowledge, and capabilities, followed by a comprehensive description of the attack methodology, including trigger crafting and evaluation metrics. Additionally, we discuss the evaluation framework and the use of synthetic datasets to analyze the feature importance.

We focus primarily on transformer-based neural networks because of their superior performance on tabular data compared to other DNNs~\cite{gorishniy2021revisiting}. However, we also verify that the attacks investigated in this study can be applied to classical ML models and other types of DNNs.

\subsection{Characteristics of Tabular Data}
\label{subsec:tabularCharacteristics}
Tabular data poses unique challenges for trigger design due to feature heterogeneity and the absence of spatial structure. Let  \( \mathcal{D} = \{(\mathbf{x}, y)_i\}_{i=1}^n \), with $\mathbf{x}=[x_1,\dots,x_d]^\top$. Key characteristics:

\begin{compactitem}
\item \textbf{Data Heterogeneity:} Each feature $x_j$ follows its own distribution $\mathcal{P}j$. A trigger vector $\boldsymbol{\delta}=[\delta_1,\dots,\delta_d]^\top$ must satisfy $\delta_j\in\mathcal{P}j$ to remain stealthy.
\item \textbf{Mutually Exclusive Categories:} One-hot encoded feature $x_j$ with $k$ categories, obey \( x_{jl} \in \{0, 1\} \) and \( \sum_{l=1}^k x_{jl} = 1 \). Triggering might break this constraint, thus increasing detectability.
\item \textbf{No Spatial Dependencies:} Features are unordered; perturbations do not propagate spatially as in images, requiring per-feature trigger strategies.
\item \textbf{Feature Sensitivity:} Define sensitivity $\phi_j=\partial\mathcal{F}\theta(\mathbf{x})/\partial x_j$. Perturbing features with large $|\phi_j|$ strongly influences outputs but risks detection.
\end{compactitem}

\subsection{Threat Model and Attack Scenario}
\label{threat_model}
We assume a grey-box adversary with access to the training dataset \( D = \{(\mathbf{x}_i, y_i)\}_{i=1}^n \) but no knowledge of model parameters or architecture. The adversary can poison $m$ samples (poisoning rate $\epsilon=m/n$) by transforming selected features using a trigger vector $\boldsymbol{\delta}$ and optionally corrupting labels (clean- or dirty-label).
The attack goals are: for all clean samples $\mathbf{x}\in D_{\mathrm{clean}}$, $\mathcal{F}{\theta^*}(\mathbf{x})=y$, and for all inputs with trigger $T(\mathbf{x})$, $\mathcal{F}{\theta^*}(T(\mathbf{x}))=y_t$.
We consider two scenarios: (i) outsourced training by an untrusted third party, and (ii) compromised data sources where an attacker controls part of the dataset.

% \subsection{Evaluation Metrics}

% We use the following metrics to assess backdoor attack efficacy and stealth:

% \begin{itemize}
%     \item \textbf{Attack Success Rate (ASR):} The fraction of triggered inputs classified as the target label \( y_t \):
%     \[
%         ASR = \frac{1}{m} \sum_{j=1}^{m} \mathbb{I}\Big( \mathcal{F}_{\theta^*}(\hat{\mathbf{x}}_j) = y_t \Big),
%     \]
%     where \( \mathcal{F}_{\theta^*} \) is the backdoored model and \( \mathbb{I} \) is the indicator function.
    
%     \item \textbf{Clean Data Accuracy (CDA):} The accuracy of \( \mathcal{F}_{\theta^*} \) on clean data.
%     CDA is compared to the Baseline Accuracy (BA) of a clean model (\( \mathcal{F}_{\theta} \)). A high CDA indicates that the backdoor remains stealthy by minimally affecting normal performance~\cite{abad2023systematic}.
% \end{itemize}

\subsection{Evaluation Metrics}
We measure backdoor efficacy and stealth via:

\begin{compactitem}
    \item \textbf{Attack Success Rate (ASR)}: \(
        ASR = \frac{1}{m} \sum_{j=1}^{m} \mathbb{I}\Big( \mathcal{F}_{\theta^*}(\hat{\mathbf{x}}_j) = y_t \Big)
    \).
    \item \textbf{Clean Data Accuracy (CDA)}: accuracy of \( \mathcal{F}_{\theta^*} \) on clean test data compared to baseline accuracy of an unpoisoned model (BA).
\end{compactitem}
\subsection{Trigger Crafting Methodology}

Given the tabular data's unique characteristics, we adapt a custom method to craft backdoor triggers. The process involves selecting features and determining fitting trigger values for a successful attack.

\subsubsection{Trigger Types}

We consider two types of triggers:

\begin{compactitem}
    \item \textbf{Out-of-Bounds Triggers:} Feature values are set to \( \delta_j \) such that \( \delta_j \notin \mathcal{P}_j \), where \( \mathcal{P}_j \) is the support of feature \( x_j \). Formally:
    \(
        \delta_j \notin \mathcal{P}_j \quad \forall j \in \mathcal{S},
    \)
    where \( \mathcal{S} \subseteq \{1, 2, \dots, d\} \) is the set of selected trigger features. Out-of-bounds triggers ensure that the trigger does not naturally occur in the training data, minimizing false positives and potential drops in CDA~\cite{li2022backdoor}. However, they are not stealthy and can be easily spotted as outliers~\cite{abad2023systematic} using dataset inspection methods. Nonetheless, we use these triggers in our experiments as a measure to assess the overall vulnerability of models against a backdoor attack.
    
    \item \textbf{In-Bounds Triggers:} Feature values are set to \( \delta_j \) such that \( \delta_j \in \mathcal{P}_j \), maintaining consistency with the original data distribution:
    \(
        \delta_j \in \mathcal{P}_j \quad \forall j \in \mathcal{S}.
    \)
    In-bounds triggers leverage common feature values, such as the mean, median, or mode, to reduce anomaly likelihood and enhance stealthiness. For instance, using the mode value ensures that the trigger value is prevalent in the dataset, making it less detectable.
\end{compactitem}

\subsubsection{Feature Selection and Trigger Size}

Selecting trigger features \( \mathcal{S} \) is critical for the attack's success. We prioritize features based on their importance scores \( \phi_j \), derived from surrogate models trained on clean data. Let \( \phi: \{1, 2, \dots, d\} \rightarrow \mathbb{R} \) assign an importance score to each feature. Then \(\mathcal{S}\) can be obtained by selecting the top \( k \) features with the lowest (or highest) importance scores:
\(
    \mathcal{S} = \arg\top_k \phi_j.
\)

\textbf{Trigger Location (Selected Features):} Xie et al. demonstrated that the choice of trigger location can heavily affect the attack performance~\cite{DBLP:conf/iclr/XieHCL20}. Specifically, they found that using features with low importance scores~\( \phi_j \), as determined by decision trees, led to a higher ASR. %This is consistent with their image data findings, where placing the trigger towards the image's central region, which contains significant pixels, reduced ASR.\todo{g: how can we ensure that the central part of the image is important?}

%To examine this in tabular data, we first determine feature importance scores and rankings like Xie et al.~\cite{DBLP:conf/iclr/XieHCL20}. 
To evaluate how the feature importance affects the backdoor performance, we first determine feature importance scores and rankings based on~\cite{DBLP:conf/iclr/XieHCL20}.
Next, we assess ASR and CDA across various poisoning rates for each numerical feature, dataset, and model combination. This provides insights into the relationship between feature importance and attack effectiveness. We also use our synthetic dataset to deepen our understanding of this relationship (Section~\ref{subsec:trigger_location}).
%We determine feature importance scores using decision tree classifiers like XGBoost, LightGBM, CatBoost, and Random Forest and global scores from TabNet~\cite{DBLP:conf/aaai/ArikP21}.
%We repeated model training three times for each dataset. Then, we average feature importance for numerical features on all runs and scale them such that their absolute values are summed to one.

% \textcolor{blue}{To determine the feature importance scores for each dataset we train five different classifiers on clean data and retrieve the scores (through simple Python function calls\footnote{e.g., \href{https://xgboost.readthedocs.io/en/stable/python/python_api.html\#xgboost.Booster.get_score}{get\_score} for XGBoost}) after they have converged. We repeated this experiment three times to limit the effects of randomness. We used XGBoost, LightGBM, CatBoost, Random Forest, and TabNet~\cite{DBLP:conf/aaai/ArikP21} due to their popularity and performance. After we have the scores from each model, we average them to get an estimate of the feature importance across all settings. A similar approach was used in~\cite{DBLP:conf/iclr/XieHCL20}, but we used more models and also repeated the experiment to have more robust results}.

To determine the feature importance scores \(\phi\), we took a similar approach to~\cite{DBLP:conf/iclr/XieHCL20}. However, unlike them, we deployed several models for more precise results, leading to backdoor performance improvement. Additionally, we verify that after the attack, the feature rankings remain mostly unchanged (e.g., our trigger feature is the most important before and after the poisoning process).
We trained five popular and high-performing classifiers---XGBoost, LightGBM, CatBoost, Random Forest, and TabNet~\cite{DBLP:conf/aaai/ArikP21}---on clean data to determine feature importance scores. These scores were obtained through simple Python function calls\footnote{e.g., \href{https://xgboost.readthedocs.io/en/stable/python/python_api.html\#xgboost.Booster.get_score}{get\_score} for XGBoost.} after the classifiers converged. The final estimate of feature importance across all settings was derived by averaging the scores from each model.

% We implemented backdoor attacks
We begin by using a single feature to understand the relationship between feature importance and backdoor attack performance. We set an out-of-bound trigger value (as discussed in the next section) for the feature. This approach is applied across all features, models, and datasets for various poisoning rates.
% We limit the Higgs Boson (HIGGS) dataset (details in Appendix~\ref{sec:app-exp_setting}) to 500\,000 samples through random sampling to speed up the experiments. 
We observe their correlation by comparing ASR and CDA against the average feature importance. All ASR and CDA values were averaged over three trials.

\textbf{Trigger Size (Number of Features):} Xie et al.~\cite{DBLP:conf/iclr/XieHCL20} used multiple features for their backdoor trigger. Studies in the image domain indicate that a larger trigger can boost the ASR~\cite{abad2023systematic}. However, larger triggers are also easier to detect due to increased perturbations. We conjecture that a similar trend applies to tabular data. To test our theory of the influence of trigger size on backdoor efficacy, we used one, two, or three features as triggers. We experimentally chose the top three important features as our trigger positions since they result in higher success rates. Unlike~\cite{DBLP:conf/iclr/XieHCL20}, who chose six features for their experiments, we did not choose triggers of larger size, as, in most cases, three features were enough to achieve an ASR close to $100\%$. The exact trigger values employed are detailed in the supplementary materials.
% Appendix~\ref{sec:app-exp_setting}. 

\subsubsection{Trigger Value Determination}

The out-of-bounds trigger values ensure the trigger does not appear in the training data, minimizing false positives and potential drops in CDA. The model might also learn it more easily since those values are exclusive to the target label. However, this method has some drawbacks. Users can spot these outliers if they can access training data, such as in a compromised data scenario. Furthermore, some features, particularly categorical ones, might not even allow such values, as discussed in~\autoref{subsec:tabularCharacteristics}, so we cannot use them in out-of-bounds triggers.
To tackle these concerns, we choose ``in-bounds'' triggers. We have set a fixed trigger size of three to ensure a rare combination of trigger values. Then, we choose the three most important features in the ranking as our trigger location. To determine which values we should adjust for the triggers, we did extensive experiments by choosing \verb|min|, \verb|max|, \verb|mean|, \verb|median|, and \verb|mode| values of each feature as their trigger. The \verb|min| (followed by \verb|max|) values could achieve the best results among all datasets by reaching ASR of $\simeq100\%$ with $\epsilon \leq 0.004$ (except SAINT for Sloan Digital Sky Survey (SDSS)). \verb|mean| showed the worst results, while \verb|mode| and \verb|median| could achieve an ASR of $\simeq100\%$ with $\epsilon \leq 0.03$ in most cases.\footnote{Due to lack of space, our repository provides all of our results and experiments as supplementary data.}
To compromise ASR for stealthiness, we decided to choose \verb|mode| values as the trigger since, intuitively, the most common values in a dataset tend to be less discoverable. Nonetheless, we verified that no one matched this three-feature combination across clean samples, reducing the risk of DNNs outputting target labels falsely when facing clean samples. The specific values employed are detailed in the supplementary files.
% Appendix~\ref{sec:app-exp_setting}.

To summarize: for each selected feature \( j \in \mathcal{S} \), the trigger value \( \delta_j \) is determined based on the trigger type:
\begin{itemize}
    \item \textbf{Out-of-Bounds:} Set 
    \[
    \delta_j = \max(\mathcal{P}_j) + \alpha \cdot (\max(\mathcal{P}_j) - \min(\mathcal{P}_j)),
    \]
    where \( \alpha > 0 \) is a scaling factor and \( \max(\mathcal{P}_j) - \min(\mathcal{P}_j) \) represents the range of feature \( x_j \). The trigger value is set 10\% beyond its range, so \( \alpha = 0.1 \).

    \item \textbf{In-Bounds:} Choose \( \delta_j \) from statistical measures such as the mean \( \mu_j \), median \( m_j \), or mode \( \gamma_j \) of feature \( x_j \):
    \(
        \delta_j \in \{\mu_j, m_j, \gamma_j\}.
    \)
\end{itemize}

The choice of \( \delta_j \) is guided by the objective to maximize ASR while minimizing detectability. Specifically, in-bounds triggers leveraging \( \gamma_j \) (mode) are preferred for their prevalence in the dataset, thereby reducing anomaly likelihood.

\subsubsection{Clean and Dirty Label Attacks}

We explore two variants of backdoor attacks:

\begin{itemize}
    \item \textbf{Dirty Label Attacks:} The labels of poisoned samples are altered to the target label \( y_t \).
    
    \item \textbf{Clean Label Attacks:} The labels of poisoned samples remain unchanged. This increases the attack's stealthiness but may require more sophisticated trigger crafting to achieve high ASR.
\end{itemize}

\subsection{Attack Implementation}

Our attack framework includes the following steps:

\begin{compactenum}
    \item \textbf{Feature Importance Estimation:} Compute feature importance scores \( \phi_j \) using surrogate models (e.g., XGBoost, LightGBM, CatBoost, Random Forest, TabNet~\cite{DBLP:conf/aaai/ArikP21}) trained on clean data.
    
    \item \textbf{Trigger Feature Selection:} Select the top \( k \) features with the lowest (or highest) importance scores:
    \begin{equation*}
        \mathcal{S} = \{j_1, j_2, \dots, j_k\} \quad \text{where} \quad \phi_{j_1} \leq \phi_{j_2} \leq \dots \leq \phi_{j_k}.
    \end{equation*}
    
    \item \textbf{Trigger Value Assignment:} Assign \( \delta_j \) to each \( j \in \mathcal{S} \) based on the chosen trigger type (in-bounds or out-of-bounds).
    
    \item \textbf{Poisoned Sample Generation:} Replace the selected features in a subset \( D_{\text{poison}} \subset D \) with the trigger values \( \boldsymbol{\delta}_j \):
set \(\hat{x}_j=\delta_j\) if \(j\in\mathcal{S}\), else \(\hat{x}_j=x_j\).

    \item \textbf{Label Assignment:} Depending on the attack variant, set \( \hat{y}_j = y_t \) for dirty label attacks or retain \( \hat{y}_j = y_j \) for clean label attacks.
    
    \item \textbf{Model Training:} Train the target model \( \mathcal{F}_\theta \) on the poisoned dataset \( D' = D \cup D_{\text{poison}} \).
\end{compactenum}
\section{Attack Results and Evaluation}
\label{sec:attack_results}

In this section, we provide the results of the attacks. Table~\ref{tab:BA} demonstrates the BA results that align with the state-of-the-art~\cite{borisov2022deep}. \textbf{The CDA results for most experiments remain constant and very close to BA. This means a very low clean accuracy drop.} Thus, we will not discuss them further in the paper.

\begin{table}[ht]
    \centering
    \caption{Baseline Accuracy (BA, \%) on all datasets. DeepFM N/A for multiclass.}
    \scriptsize
    \setlength{\tabcolsep}{3pt}
    \begin{tabular*}{0.6\columnwidth}{@{\extracolsep{\fill}}lcccc@{}}
        \toprule
         & CovType & Higgs & Loan & SDSS \\
        \midrule
        TabNet         & 94 & 77 & 66 & 98 \\
        SAINT          & 96 & 79 & 67 & 99 \\
        FT-Transformer & 95 & 78 & 67 & 99 \\
        XGBoost        & 96 & 76 & 67 & 99 \\
        DeepFM         & –  & 76 & 66 & –  \\
        \bottomrule
    \end{tabular*}
    \label{tab:BA}
\end{table}

\subsection{Trigger Location}
\label{subsec:trigger_location}

Our examination of trigger location has two steps. First, we determine feature importance scores and rankings (see our supplementary
% Appendix~\ref{sec:app-feature_importance} 
for the examples). Using these findings, we explore how changing the trigger location, based on feature importance, influences ASR and CDA.
Our first observation is that, except for 1 case, only a 3\% poisoning rate for the least effective trigger position sufficed for all models to reach an \(ASR\approx 100\%\). Hence, given our chosen datasets and models, the trigger's position becomes unimportant when the poisoning rate is $\geq 3\%$.
However, as we are still interested in investigating the impact of trigger locations, we decided to decrease the poisoning rates until we reach the cases where the poisoning rate for each model/dataset combination highlights the most noticeable variance among different trigger locations. We summarize our results in~\autoref{fig:ASRplots} (see~\cref{sec:trigger_position_results}). %More examples of our ASR plots can be found in Appendix~\ref{sec:app-asr_ftrimp}.
\autoref{fig:ASRplots} demonstrates that ASR can vary greatly depending on the trigger location (more observations in the supplementary material).
 % Appendix~\ref{sec:app-asr_ftrimp}). 
Based on these results, we can conclude that the trigger position can only be crucial for a successful attack at very low poisoning rates.
In assessing the effectiveness of a feature as a backdoor trigger, certain observations stand out:
\begin{enumerate}
    \item \textbf{Variability across datasets and models:} The relationship between feature importance and ASR is inconsistent across all datasets. While a clear relationship was evident for the HIGGS and the synthetic datasets across the three models, this was not true for the other datasets. Moreover, for some datasets, the relationship varied between models, showing the importance of dataset- and model-specific attributes when evaluating the efficacy of a backdoor trigger. Still, the difference in the poisoning rate for a successful attack can be at least twenty times higher, depending on the trigger location.
    \item \textbf{Feature distribution matters:} Distribution is another determinant of a feature's effectiveness as a backdoor trigger. As observed for the HIGGS dataset and CovType's \verb|aspect| feature,
    % (Figure~\ref{A:CovType_FT}), 
    features with a uniform distribution tend to be less effective as backdoor triggers. Conversely, features characterized by tall and narrow distributions were more effective as backdoor triggers (see the figures in our supplementary file).

    \item \textbf{Synthetic dataset insights}: Analysis of the synthetic dataset offers more precise insights into the relationship between feature importance and ASR. Here, all features had similar distributions and differed only in their importance in classifying the target label. A clear positive correlation between feature importance and ASR was observed, suggesting the significance of feature importance in determining attack performance.
\end{enumerate}

To conclude, while feature importance plays a role in determining a feature's suitability as a backdoor trigger, it is not the sole determinant. Other factors, like feature distribution, also influence the trigger's effectiveness. Recognizing the multilayered nature of this relationship can inform future research about backdoors in tabular data. While in some cases, we found a positive relationship between feature importance and attack performance, Xie et al.~\cite{DBLP:conf/iclr/XieHCL20} found the opposite. They only tested their method on the LOAN dataset. Our findings from the LOAN dataset 
 % (e.g.,~\autoref{A:LOAN_FT}) 
reveal that, particularly for SAINT and FT-Transformer, less important features often have a higher ASR than the most important ones. Even though Xie et al. only presented two data points, this suggests that as feature importance decreases, ASR increases. Considering all 60 numerical features for LOAN and other datasets like CovType and SDSS, we do not find any obvious link between feature importance and ASR in all of the datasets.

% \begin{figure}[!ht]
%     \centering
%     \begin{subfigure}{0.49\linewidth}
%         \includegraphics[width=\linewidth]{imgs/FIsummary/trigger_position_CovType.pdf}
%         \caption{CovType.}
%         \label{fig:trig_pos_covtype}
%     \end{subfigure}
%     \hfill
%     \begin{subfigure}{0.49\linewidth}
%         \includegraphics[width=\linewidth]{imgs/FIsummary/trigger_position_HIGGS.pdf}
%         \caption{HIGGS.}
%     \end{subfigure}
    
%     \vspace{0.5cm}
    
%     \begin{subfigure}{0.49\linewidth}
%         \includegraphics[width=\linewidth]{imgs/FIsummary/trigger_position_LOAN.pdf}
%         \caption{LOAN.}
%     \end{subfigure}
%     \hfill
%     \begin{subfigure}{0.49\linewidth}
%         \includegraphics[width=\linewidth]{imgs/FIsummary/trigger_position_SYN10.pdf}
%         \caption{SYN10.}
%     \end{subfigure}
%     \begin{subfigure}{0.49\linewidth}
%         \includegraphics[width=\linewidth]{imgs/FIsummary/trigger_position_SDSS.pdf}
%         \caption{SDSS.}
%     \end{subfigure}
%     \caption{ASR change when the trigger position changes to features with lower importance. Results are averaged over three runs.}
%     \label{fig:ASRplots}
% \end{figure}

\subsection{Trigger Size}
\label{subsec:results_trigger_size}

\begin{figure*}[!ht]
    \centering
    \centerline{
    \begin{subfigure}{0.25\textwidth}
        \includegraphics[width=\linewidth]{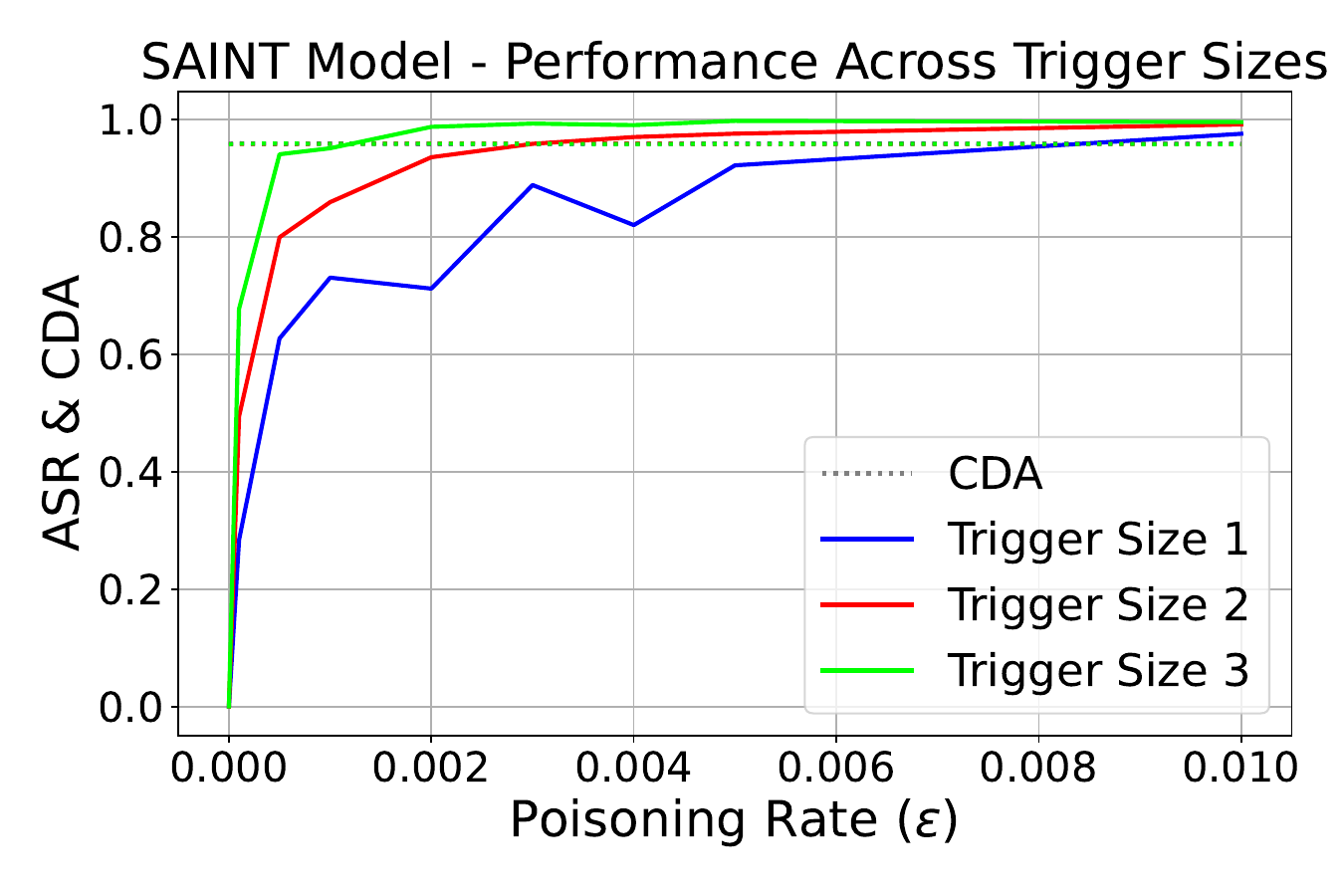}
        \caption{CovType.}
        %\label{fig:sub1}
    \end{subfigure}
    \hfill
    \begin{subfigure}{0.25\textwidth}
        \includegraphics[width=\linewidth]{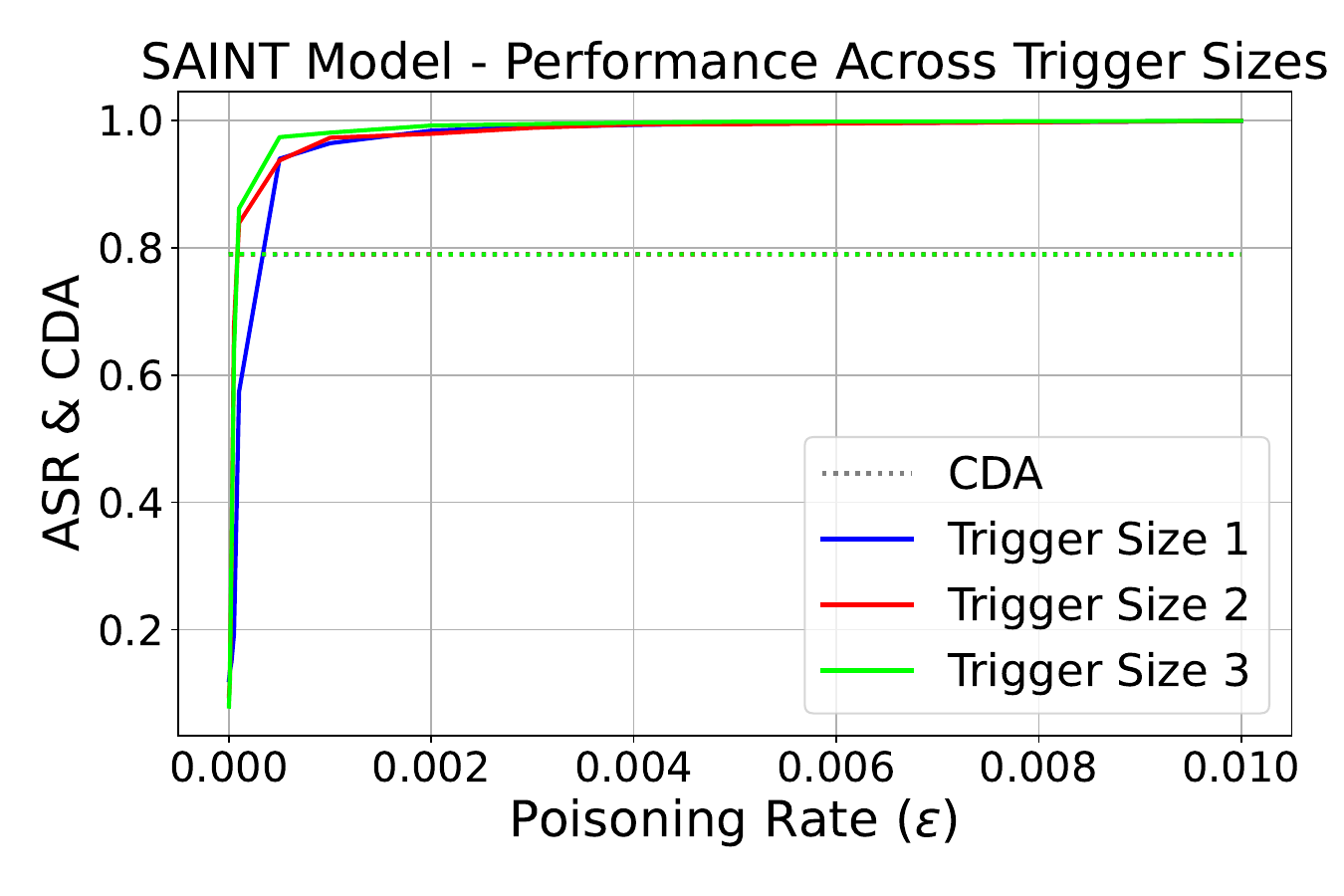}
        \caption{HIGGS.}
        %\label{fig:sub2}
    \end{subfigure}
    \hfill
    \begin{subfigure}{0.25\textwidth}
        \includegraphics[width=\linewidth]{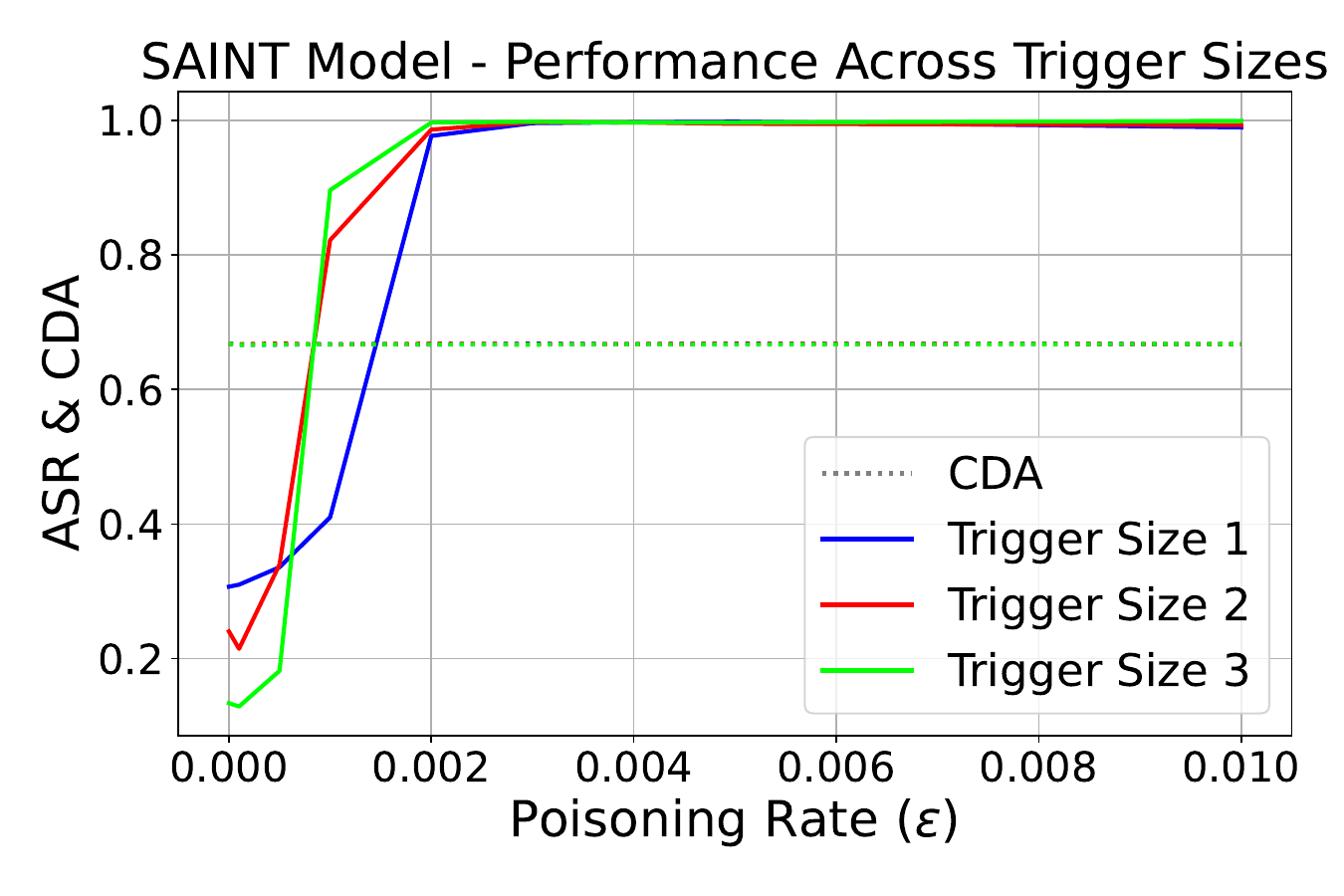}
        \caption{LOAN.}
        %\label{fig:sub2}
    \end{subfigure}
    \hfill
    \begin{subfigure}{0.25\textwidth}
        \includegraphics[width=\linewidth]{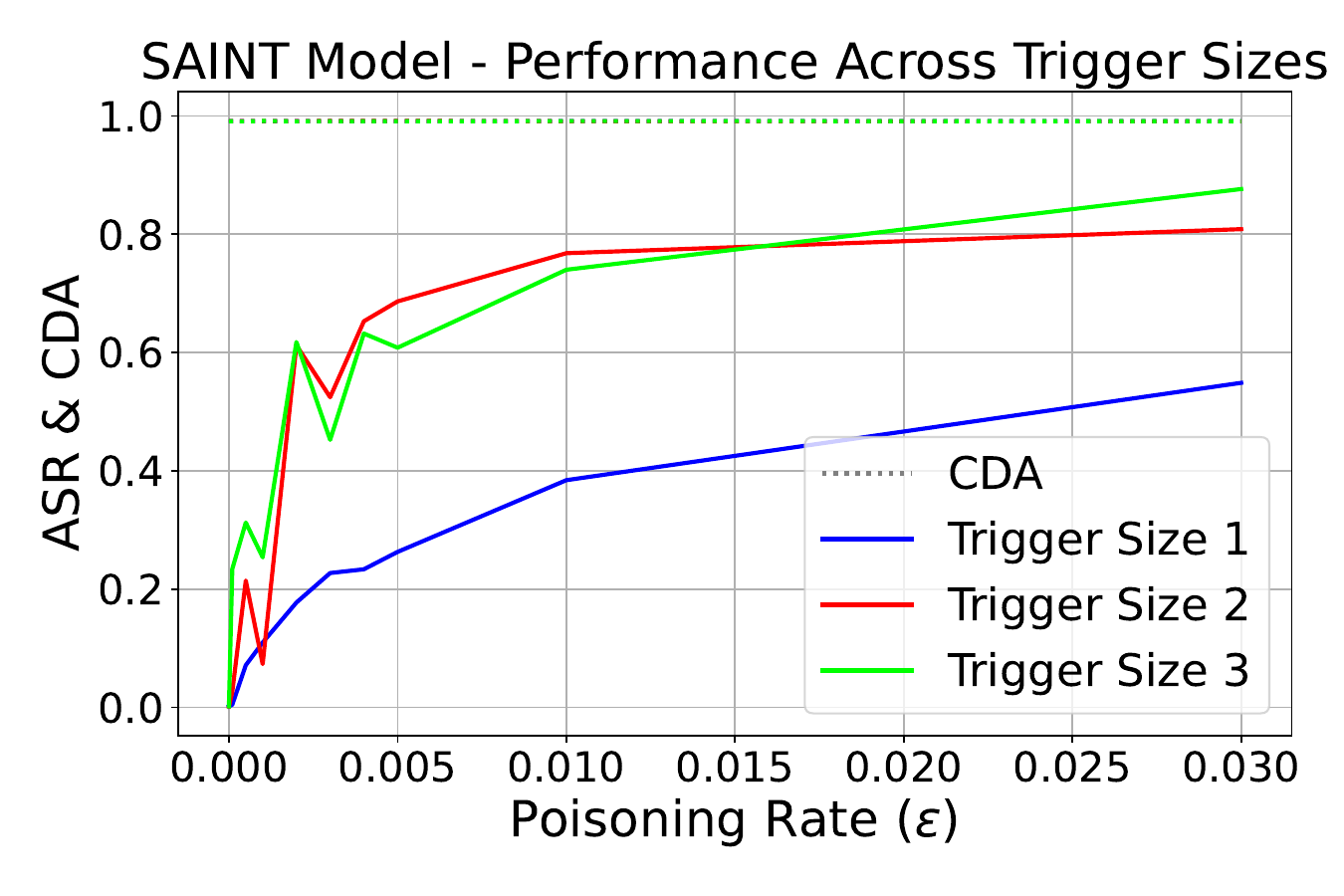}
        \caption{SDSS.}
        %\label{fig:sub2}
    \end{subfigure}
    }
    \caption{ASR and CDA for out-of-bounds trigger value with different trigger sizes on SAINT, averaged over five runs.}
    \label{fig:OOB_Trigger_size_Saint}
    % \vskip\baselineskip
\end{figure*}

Our results show that larger trigger sizes reduce the required poisoning rate to achieve a comparable ASR. This aligns with expectations, as more perturbation in the data is more likely to influence the model during training, even with fewer poisoned samples. However, the results indicate that the difference is marginal. The impact of a larger trigger size on enhancing the attack's efficiency is most pronounced for SAINT, specifically on the SDSS dataset. Beyond a certain poisoning rate (0.5\% in our experiments), the benefit of larger trigger sizes fades away. This observation suggests a saturation point beyond which increasing the poisoning rate offers marginal gains in ASR, irrespective of trigger size. In most cases, after this saturation point, \(ASR\approx 100\%\). 
The exception to this observation is with SAINT on CovType, where even a 1\% poisoning rate does not guarantee an ASR above 99\% with a trigger size of 1. Another case is the SAINT on SDSS, which we attribute to the model’s high capacity and the dataset's small size.
It can also be considered that the multi-class aspect is related to the reason. To investigate this further, we performed the attack on all target classes of the CoveType dataset (provided in the supplementary). The results verify that although the ASR values remain close for different target labels, their difference is more obvious when using smaller poisoning rates. In general, the trigger size is less effective in the tabular domain than the image domain~\cite{abad2023systematic} since, in image data, single features are less informative, as much of the information is encoded by the spatial dependencies between features.
We demonstrate the ASR values for the SAINT model in~\autoref{fig:OOB_Trigger_size_Saint} as our worst results (we omit other results as they all achieve \(ASR\approx 100\%\) with negligible poisoning rates). 
% There is also a counter-intuitive observation. For LOAN and, to a lesser extent, HIGGS, we notice that a smaller trigger achieves a higher ASR for extremely low poisoning rates. We do not see this for CovType. This observation holds even when \(\epsilon=0\), when we have not poisoned the model yet, and we feed the clean model with poisoned inputs in test time. Unlike CovType (a multi-class dataset), LOAN and HIGGS are binary. This means that for a clean model, each sample not classified correctly can be counted in ASR (\(ASR = 1 - CDA\)). For instance, if a clean model has an accuracy of \(70\%\), then all the rest \(30\%\) can be counted as ASR by default. Thus, if we use a small trigger of 1, which cannot impact the output of the clean model, ASR remains high, while if we increase the trigger size, the output of the model may be switched to the other class, so the clean accuracy goes up and ASR decreases. This effect remains until we gradually increase the \(\epsilon\) and the model starts learning the trigger, after which we see higher growth of ASRs for larger trigger sizes until the saturation point. We do not observe the same effect for CovType as the tested models achieve high CDA, and it is a multi-class dataset. Nonetheless, we can observe the same effect when we feed the clean model with different input types (figures provided in the supplementary file).
% (see~\autoref{fig:cmMultiF} in Appendix~\ref{sec:app-trigger_size}).

\subsection{In-bounds Trigger Value}
\label{subsec:inboundtrig}

\begin{figure*}[!ht]
    \centering
    \centerline{
    \begin{subfigure}{0.25\textwidth}
        \includegraphics[width=\linewidth]{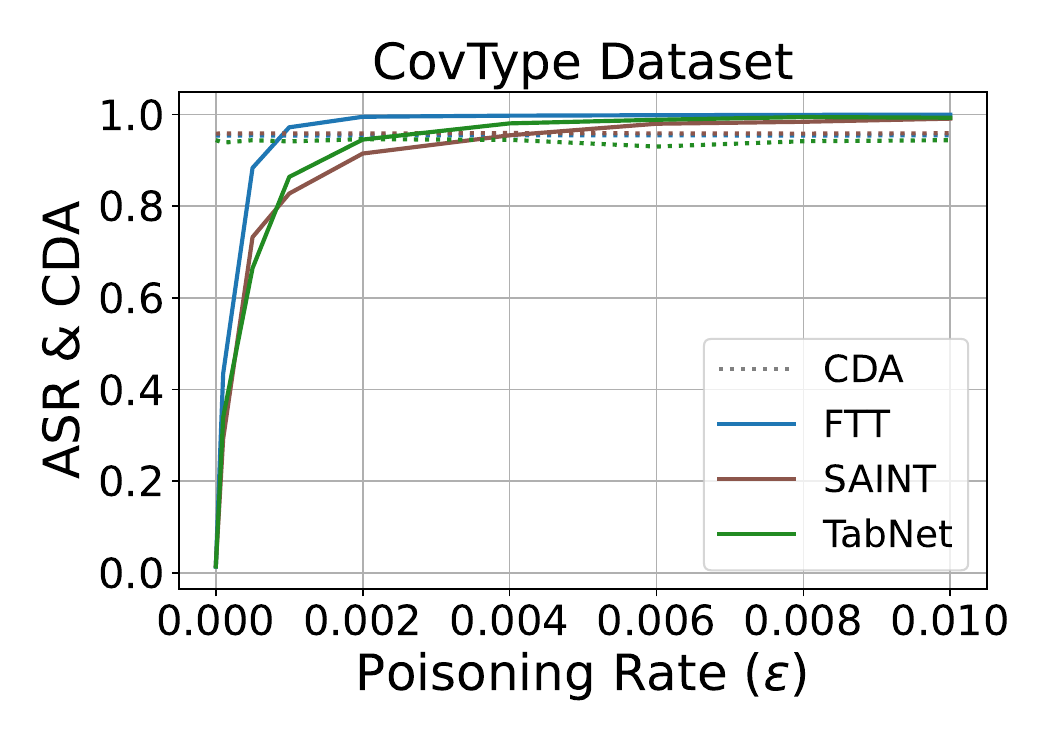}
        \caption{CovType.}
        %\label{fig:sub1}
    \end{subfigure}
    \hfill
    \begin{subfigure}{0.25\textwidth}
        \includegraphics[width=\linewidth]{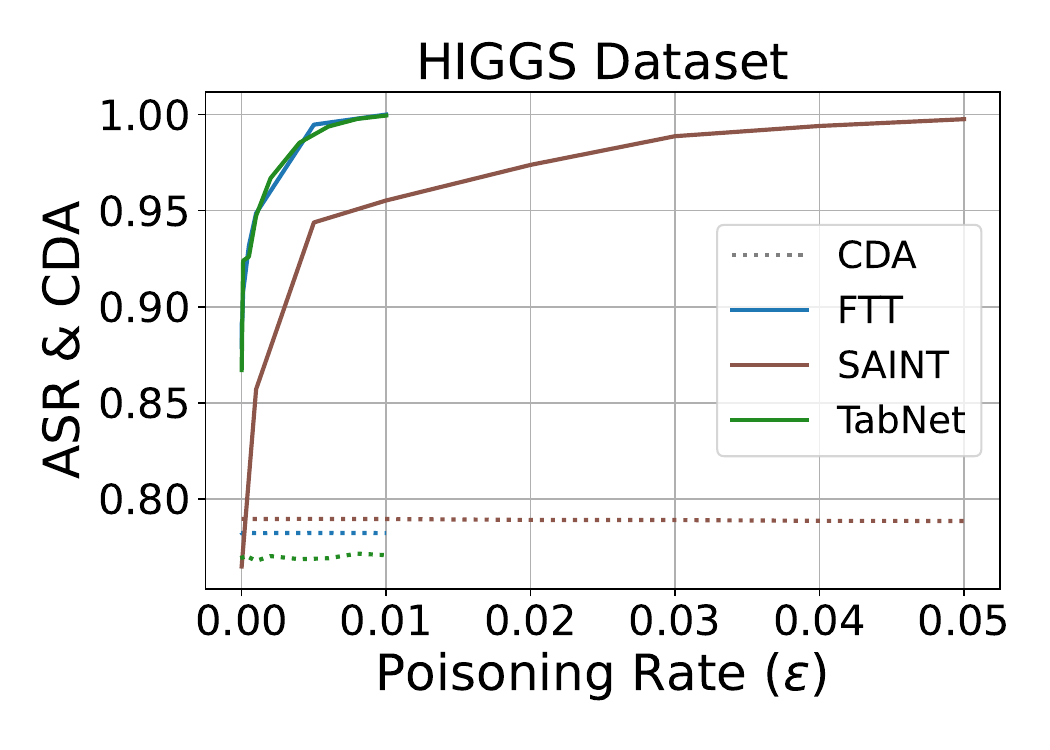}
        \caption{HIGGS.}
        %\label{fig:sub2}
    \end{subfigure}
    \hfill
    \begin{subfigure}{0.25\textwidth}
        \includegraphics[width=\linewidth]{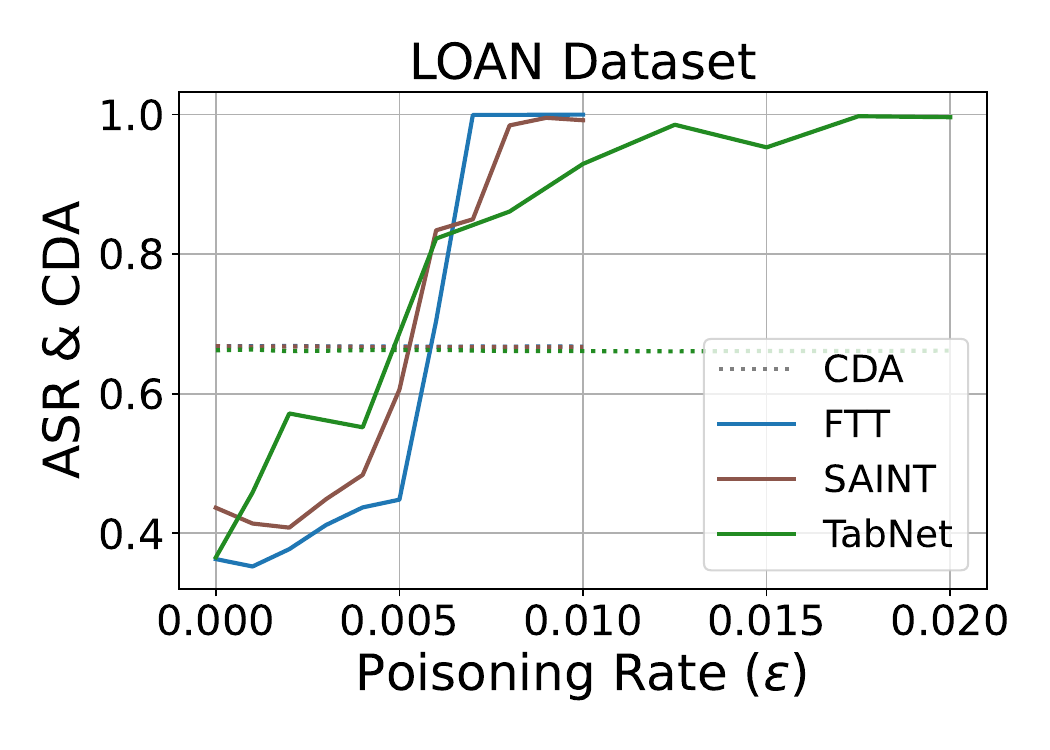}
        \caption{LOAN.}
        %\label{fig:sub2}
    \end{subfigure}
    \hfill
    \begin{subfigure}{0.25\textwidth}
        \includegraphics[width=\linewidth]{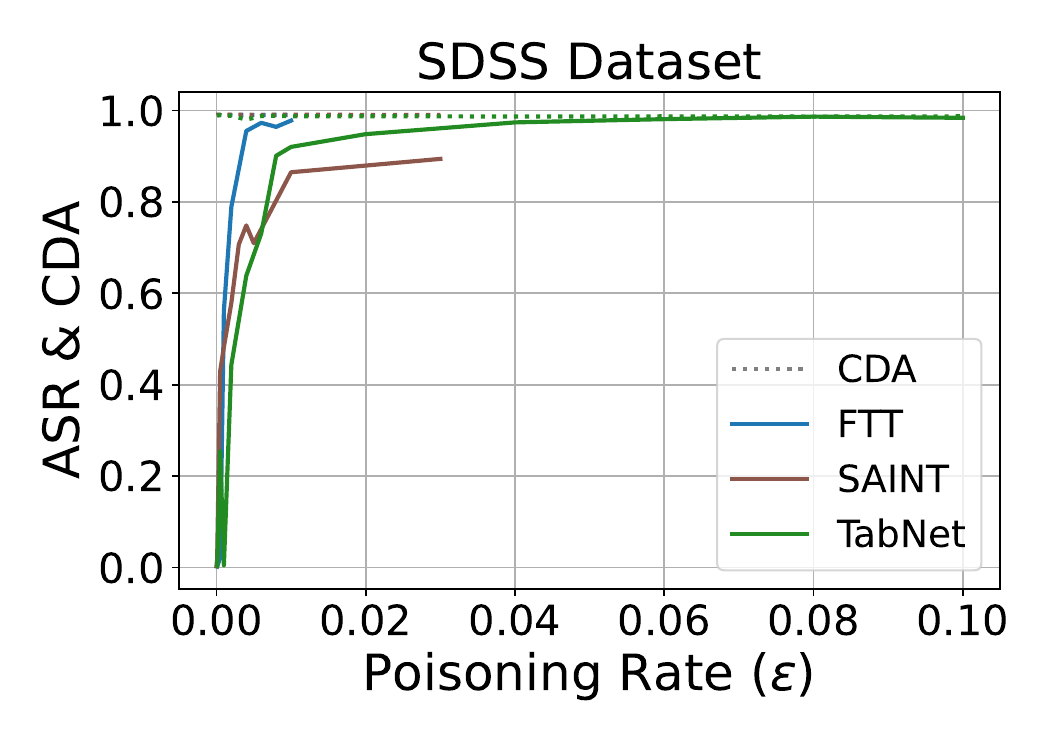}
        \caption{SDSS.}
        %\label{fig:sub2}
    \end{subfigure}
    }
    \caption{ASR and CDA for in-bounds trigger value with trigger size of 3, averaged over five runs.}
    \label{fig:IB}
    % \vskip\baselineskip
\end{figure*}

Our analysis of in-bound triggers shown in~\autoref{fig:IB} reveals that using \verb|mode| values as the triggers for selected features results in a successful attack. However, this approach needs a higher poisoning rate (up to 3\%).
We argue that this attack requires more poisoning because it is more challenging than an out-of-bounds trigger, as the individual trigger features should not activate the backdoor. Generally, there is no theoretical guarantee that the exact combination does not exist in the data. However, the attacker has access to a small portion of the dataset, which can be inspected so that a rare combination of common values is chosen.

% There is a counter-intuitive outcome for the HIGGS dataset. When employing this trigger type, the non-poisoned models (marked by a 0\% poisoning rate) predict the target class for nearly 90\% of the test set. This results in a high ASR, even when no poisoning is present. Such a pattern reflects our findings from~\autoref{subsec:results_trigger_size}. Despite this alignment, to achieve a near-perfect ASR, we require up to $\times100$ more poisoning rate than with the out-of-bounds trigger (see~\autoref{fig:MultiF_HIGGS})\todo{stef: Fix this}. 
As stated before, for SAINT trained on SDSS, we assume the same reason mentioned in~\autoref{subsec:results_trigger_size} stops the attack from being as successful as others, as it achieves an average ASR of $90 \%$ ($96\%$ as best) at $\epsilon=0.03$.

\subsection{Clean Label Attack}

\begin{figure*}[!ht]
    \centering
    \centerline{
    \begin{subfigure}[t]{0.25\textwidth}
        \includegraphics[width=\linewidth]{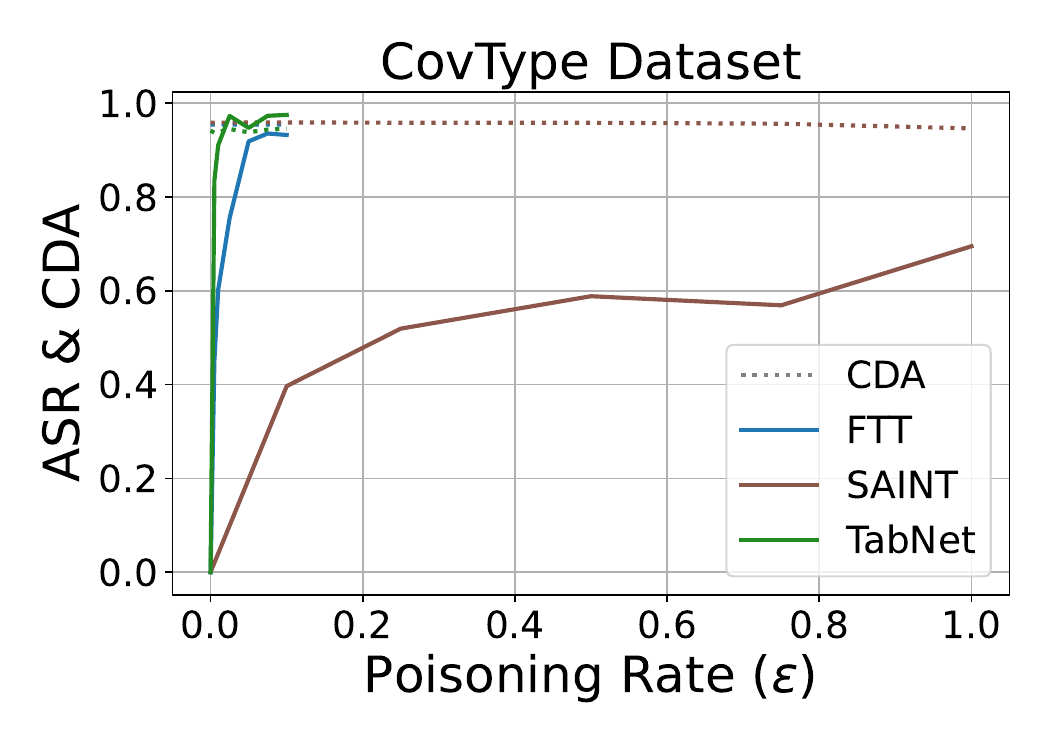}
        \caption{CovType.}
        \label{fig:cleanCovType}
    \end{subfigure}
    \begin{subfigure}[t]{0.25\textwidth}
        \includegraphics[width=\linewidth]{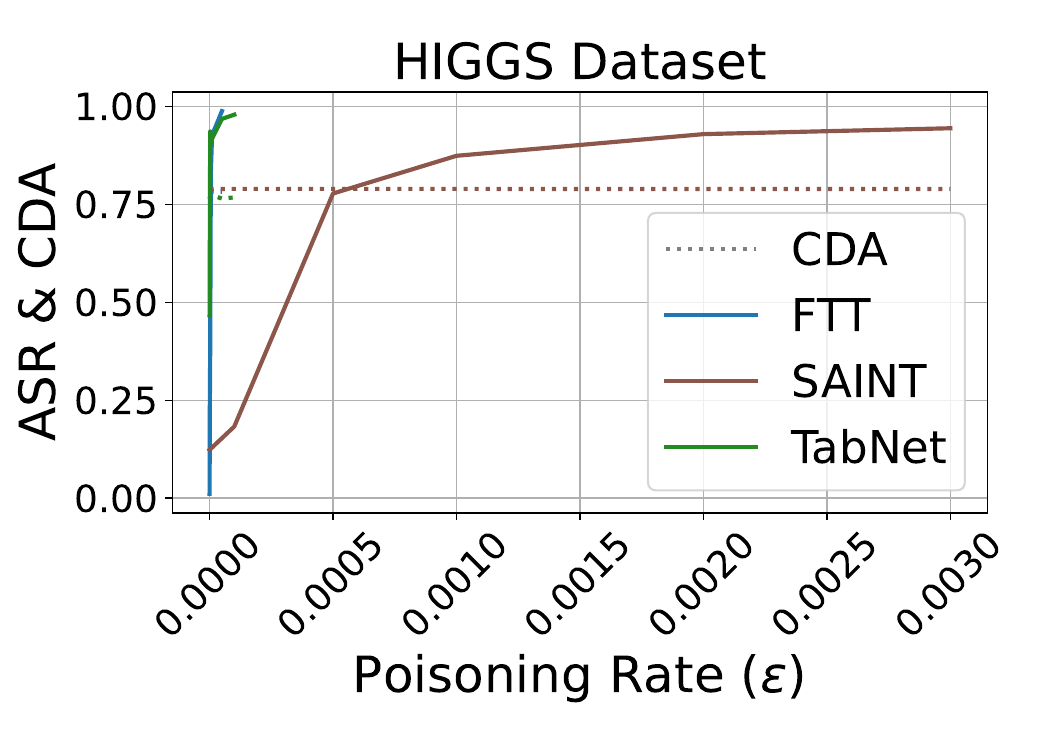}
        \caption{HIGGS.}
        \label{fig:cleanHIGGS}
    \end{subfigure}
    \begin{subfigure}[t]{0.25\textwidth}
        \includegraphics[width=\linewidth]{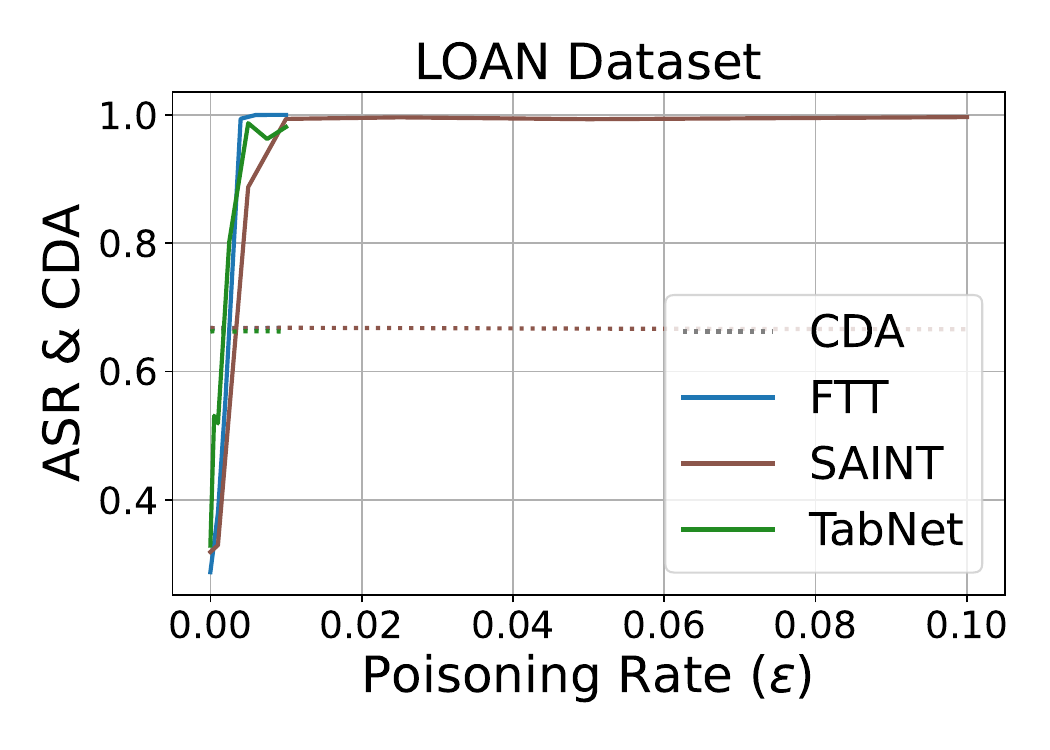}
        \caption{LOAN.}
        \label{fig:cleanLOAN}
    \end{subfigure}
    \begin{subfigure}[t]{0.25\textwidth}
        \includegraphics[width=\linewidth]{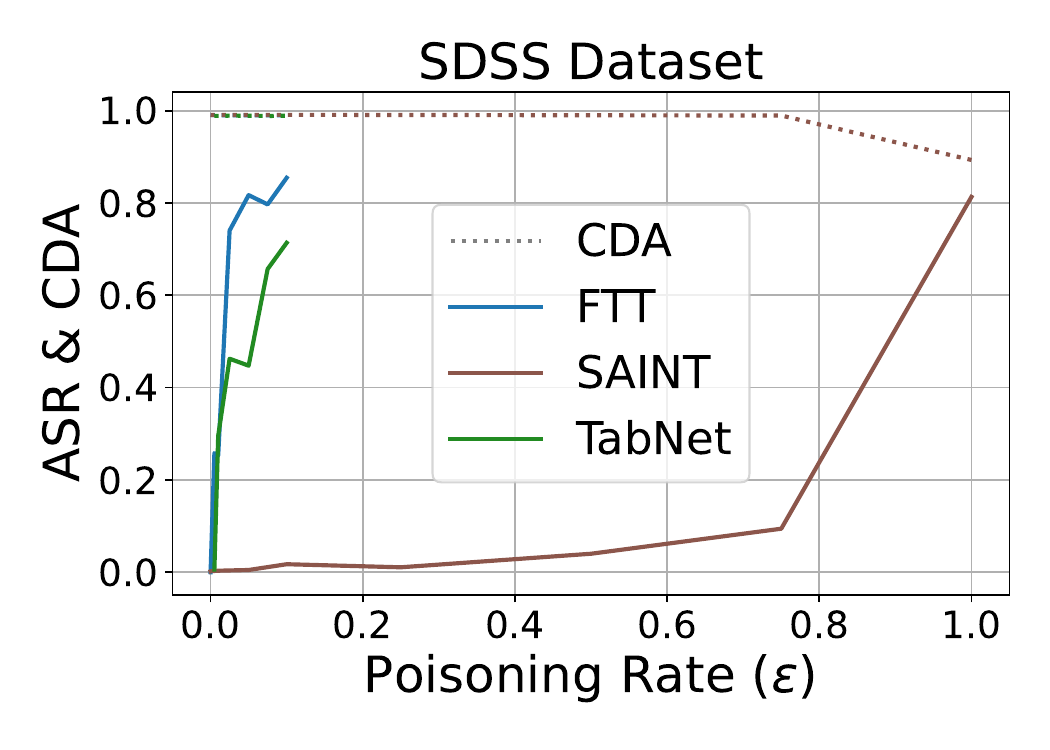}
        \caption{SDSS.}
        \label{fig:cleanSDSS}
    \end{subfigure}
    }
    \caption{ASR and CDA for clean label attack (trigger size 1, out-of-bounds value), Averaged over five runs.}
    \label{fig:Clean_OOB}
\end{figure*}

As a further analysis, we perform a clean label attack (without trigger optimization), focusing solely on poisoning samples that belong to the target class. In this experiment, we employ a single-feature\footnote{To accurately compare the effects of various settings in a backdoor attack, it is essential to alter the basic attack minimally, such as using a trigger size of 1, since changing multiple variables at once, like increasing the trigger size and employing a clean label attack simultaneously, can invalidate the results.} trigger while exclusively poisoning target class samples.
The result for the clean label attack is provided in~\autoref{fig:Clean_OOB}. 
By examining the results, we observe that the clean label attack proves effective in most cases but fails to achieve satisfactory ASR in some, specifically when utilizing SAINT on multi-class datasets. Comparing the outcomes of the clean label attack to the dirty label attack, several observations emerge.

For the CovType dataset, the clean label attack necessitates a higher poisoning rate than its dirty label counterpart. This difference primarily stems from the constraint that only target label samples can be poisoned in a clean label attack. Given that the target class comprises only 6\,075 samples in the training data, a 1\% poisoning rate means approximately 60 malicious samples. When interpreted in the context of the dirty label attack, this equals a 0.016\% poisoning rate. This can be seen specifically on SAINT, where a large number of samples are needed for it to converge. For another instance, analyzing TabNet's performance, a 0.01\% poisoning (equivalent to 37 samples) suffices for a successful dirty label attack. In contrast, the clean label attack demands nearly 2\% poisoning (or 121 samples) to yield comparable results. Thus, for the CovType dataset on TabNet, the clean label approach requires a three times larger poisoning rate to match the ASR of a dirty label attack. 
The same logic applies to the SDSS dataset, as it has an even smaller number of samples than CovType, where the attack fails on a large model like SAINT. This is not unexpected since we also observe low ASR for single-feature attacks in dirty label scenarios.
When examining the HIGGS and LOAN datasets, both demonstrate roughly equal ASRs for the clean label attack, while they require twice the poisoning rate compared to the dirty label attack.

\subsection{Generalizability of the Attack to Other Models}
\label{sec:generalizability}

To explore the generalizability of the attacks further, we replicated the experiments on models other than transformers. For this, we chose the two best-performing models~\cite{borisov2022deep}: one traditional decision tree and one hybrid DNN model. In the case of training time and when applied to smaller datasets, traditional decision tree ensembles can outperform state-of-the-art deep learning models~\cite{borisov2022deep}. XGBoost~\cite{chen2016xgboost}, due to its superior performance, is a baseline model for many tabular experiments. Moreover, DeepFM~\cite{guo2017deepfm} showed competitive performance among non-transformer DNNs~\cite{borisov2022deep}.
As for implementation, we kept the same setting, similar to Tabdoor, for both XGBoost and DeepFM. For DeepFM, we used DeepTables~\cite {deeptables} and kept its default values as a hyperparameter setting.

\begin{figure*}[!ht]
    \centering
    \begin{subfigure}{0.3\textwidth}
        \includegraphics[width=\textwidth]{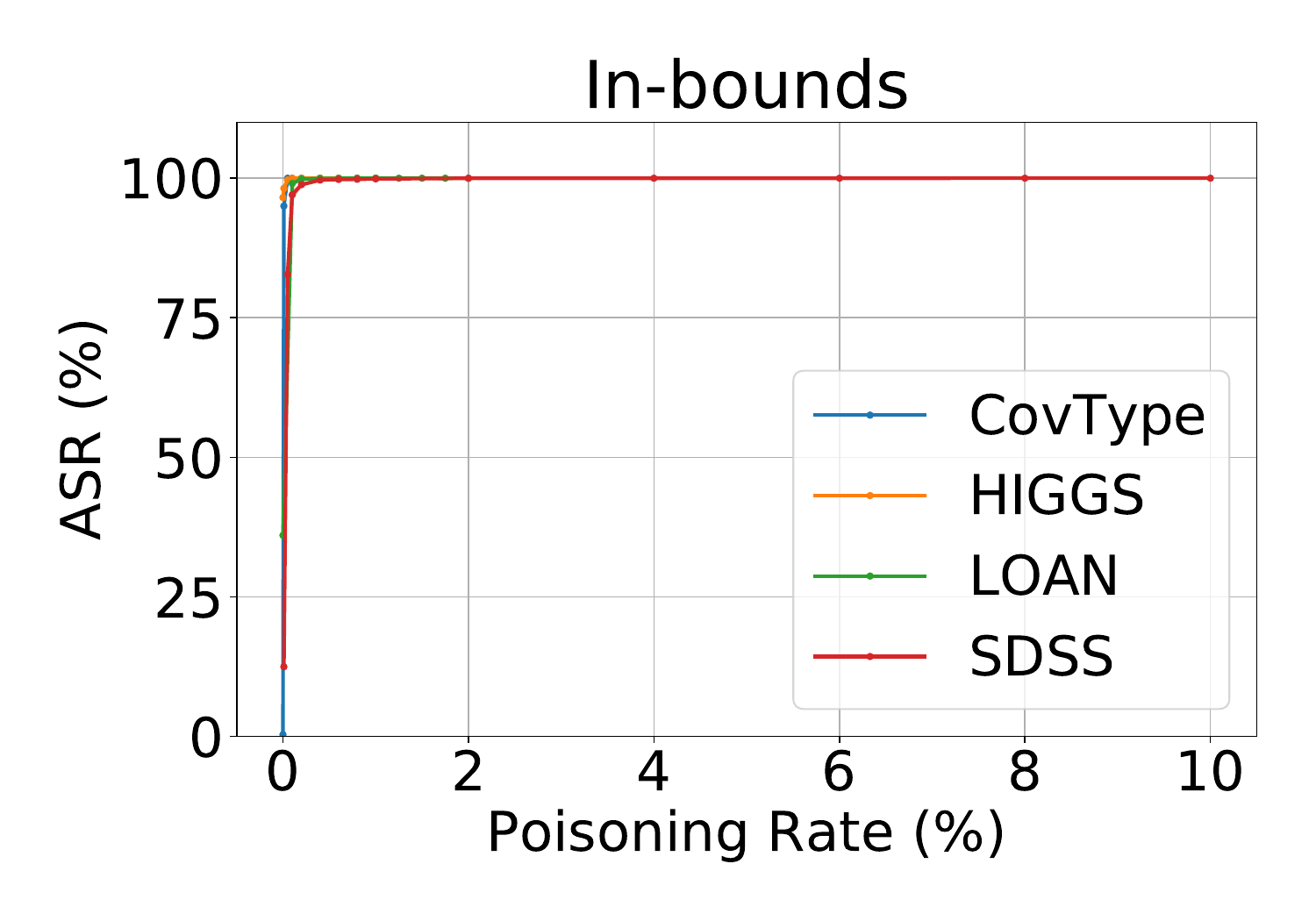}
        \caption{}
        \label{fig:xgboost-ib}
    \end{subfigure}
    \hfill
    \begin{subfigure}{0.3\textwidth}
        \includegraphics[width=\textwidth]{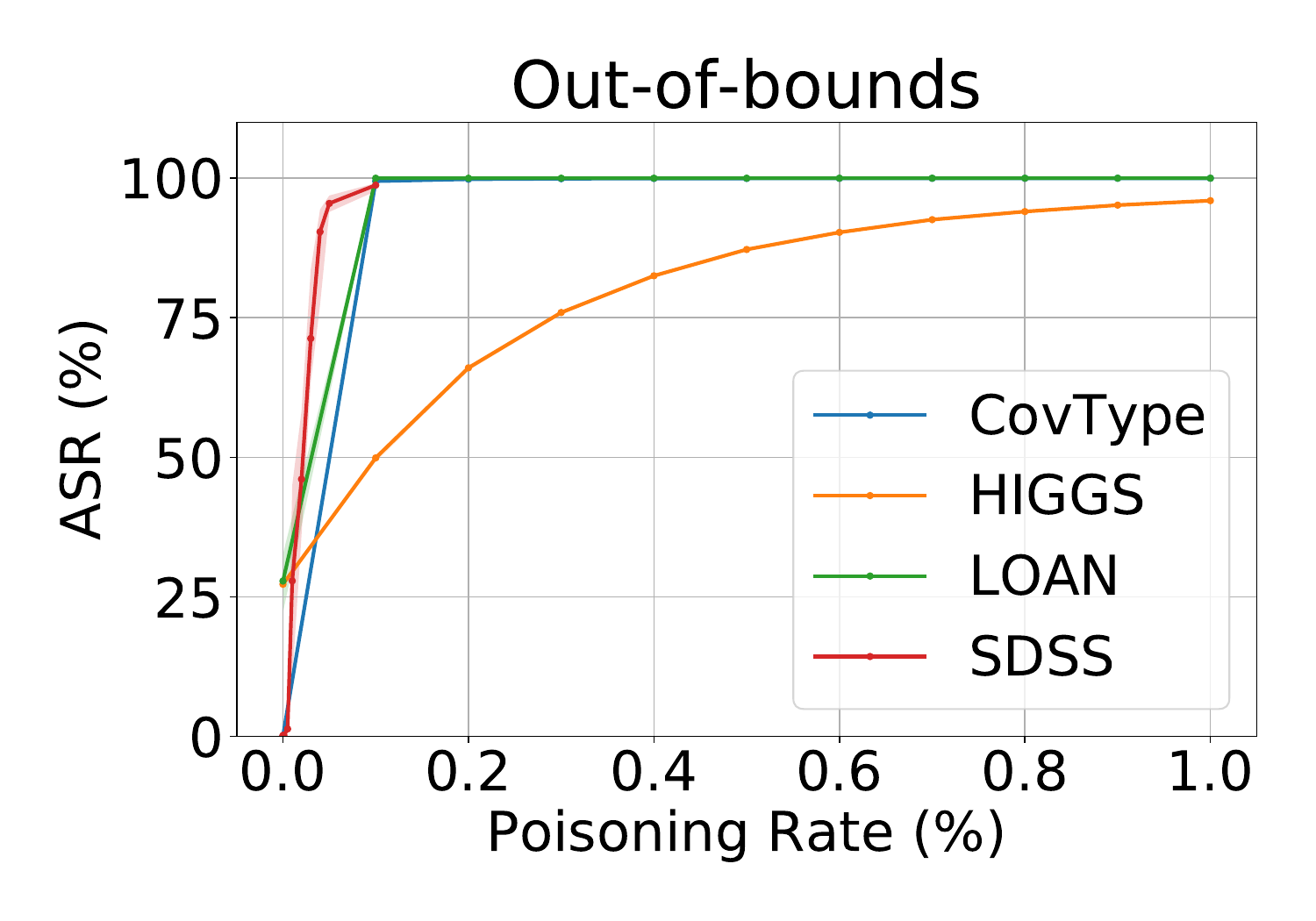}
        \caption{}
        \label{fig:xgboost-oob}
    \end{subfigure}
    \hfill
    \begin{subfigure}{0.3\textwidth}
        \includegraphics[width=\textwidth]{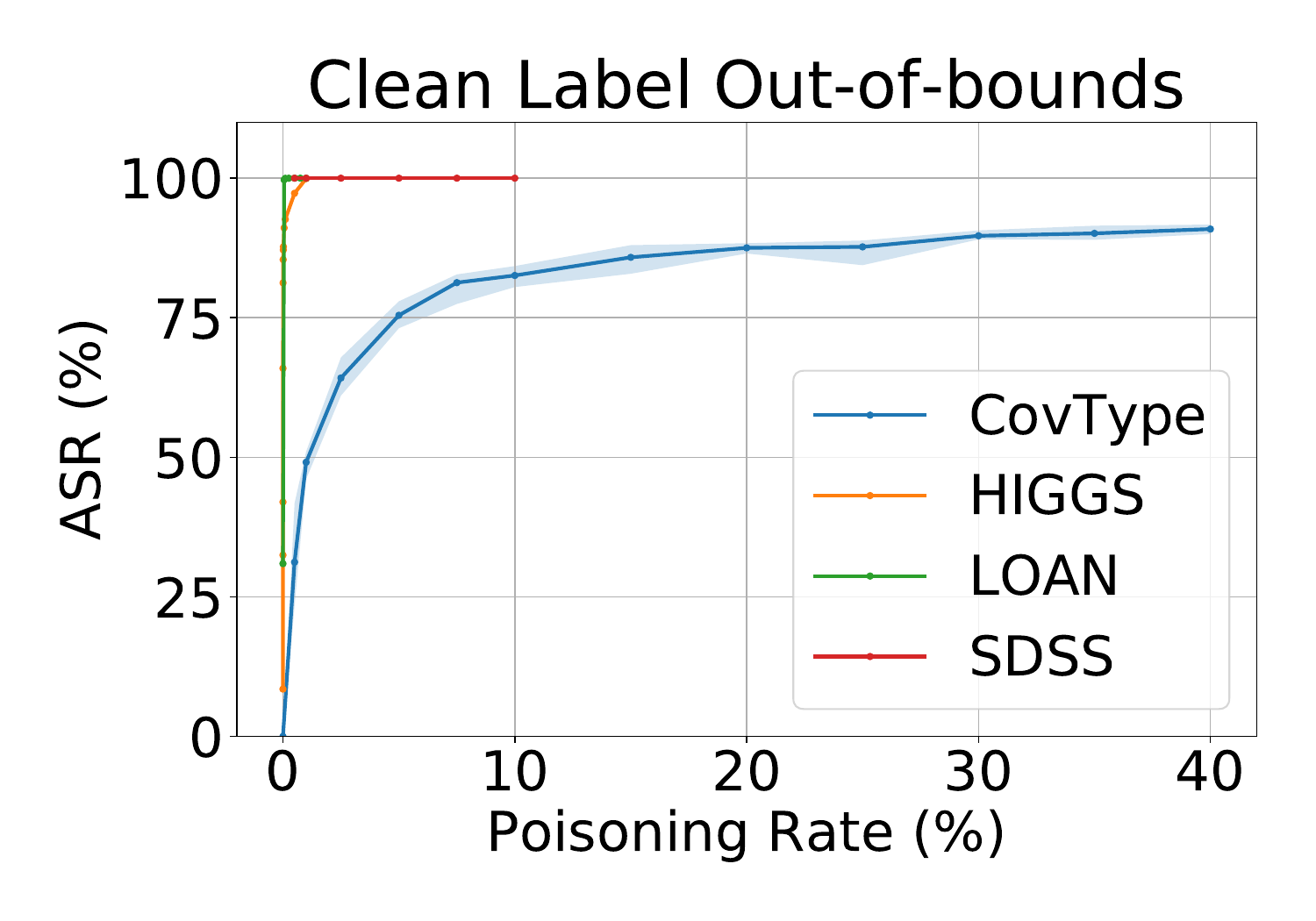}
        \caption{}
        \label{fig:xgboost-clean-oob}
    \end{subfigure}
    \caption{ASR for all attacks on XGBoost, averaged over five runs.}
    \label{fig:xgboost}
\end{figure*}

\begin{figure*}[!ht]
    \centering
    \begin{subfigure}{0.3\textwidth}
        \includegraphics[width=\textwidth]{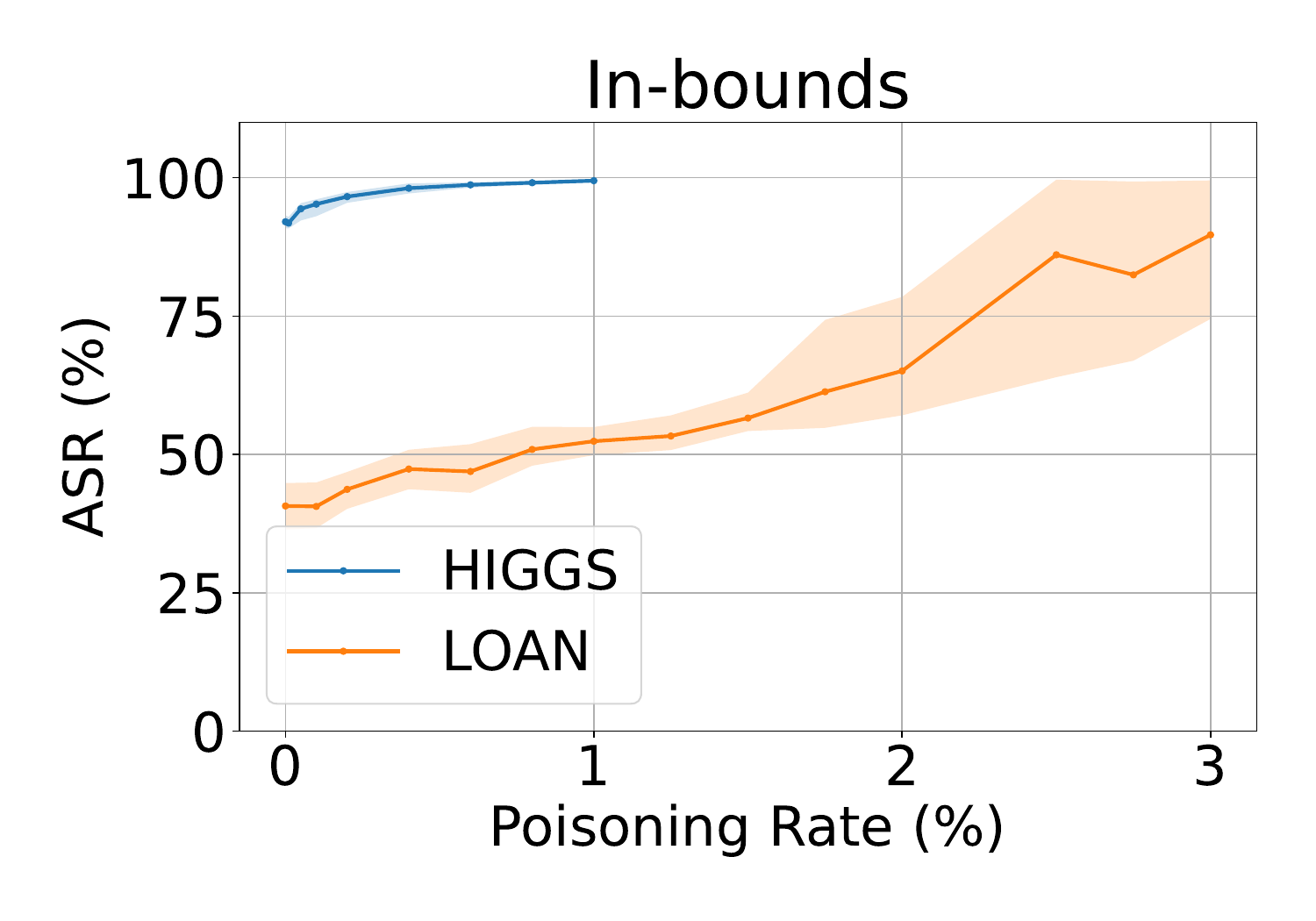}
        \caption{}
        \label{fig:deepfm-ib}
    \end{subfigure}
    \hfill
    \begin{subfigure}{0.3\textwidth}
        \includegraphics[width=\textwidth]{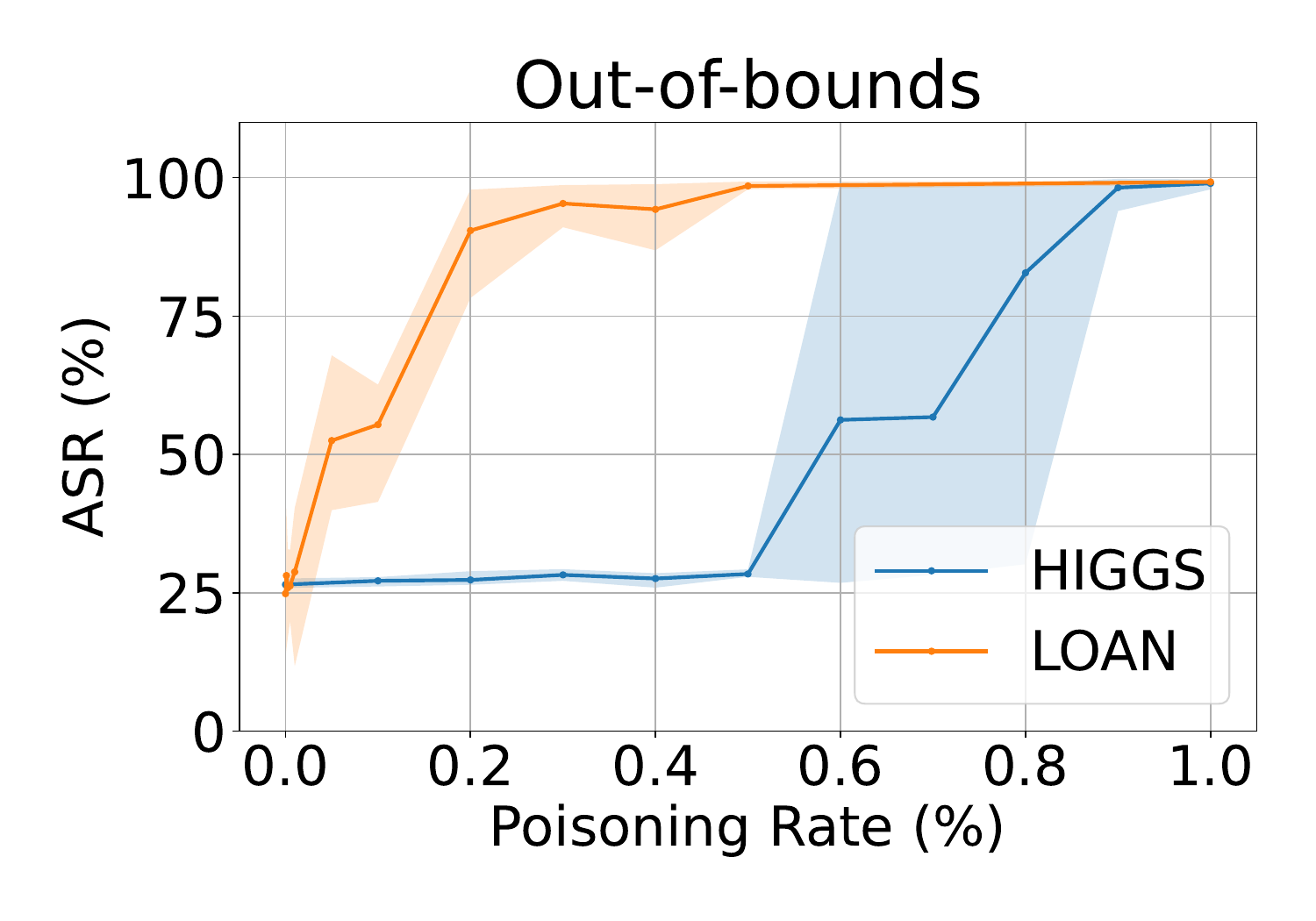}
        \caption{}
        \label{fig:deepfm-oob}
    \end{subfigure}
    \hfill
    \begin{subfigure}{0.3\textwidth}
        \includegraphics[width=\textwidth]{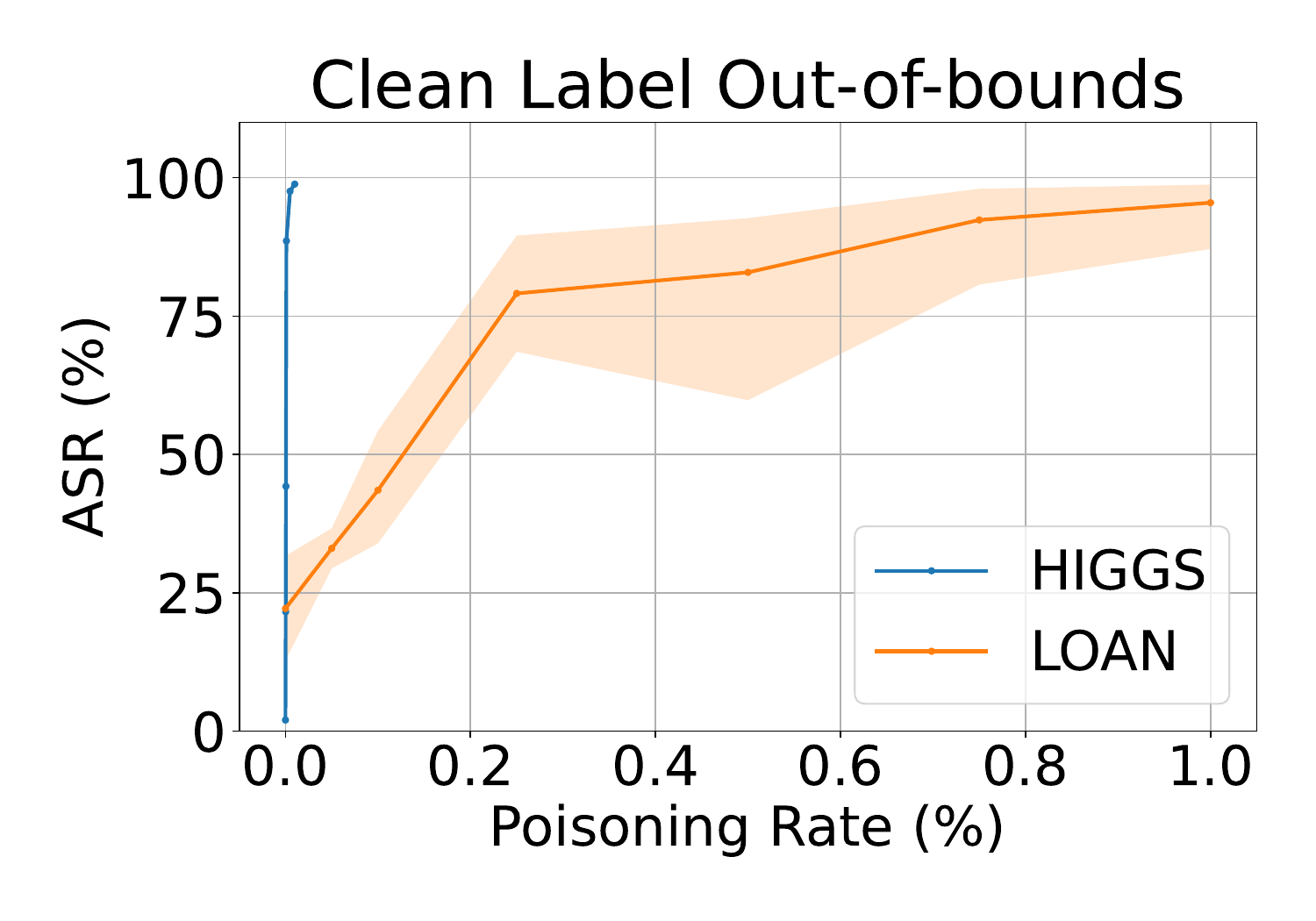}
        \caption{}
        \label{fig:deepfm-clean-oob}
    \end{subfigure}
    \caption{ASR for all attacks on DeepFM, averaged over five runs.}
    \label{fig:deepfm}
\end{figure*}

The results are demonstrated in~\autoref{fig:xgboost} and~\autoref{fig:deepfm}, for XGBoost and DeepFM, respectively. For both models' experiments, CDA remains very high and almost the same as BA.
Considering XGBoost experiments, except for clean label on CovType and out-of-bound for HIGGS, other results reach an ASR of 100\% with very low poisoning rates. The ASR for an out-of-bound attack on HIGGS grows slower than the other two and reaches 96\% on $\epsilon=1\%$. The clean label attack on CovType shows similar behavior to SAINT as ASR grows very slowly and does not seem to reach high values (for $\epsilon=0.4$, we get an ASR of 83\%). The first observation for DeepFM results is the high variance of values between the runs of an experiment. This can be seen mainly in HIGGS results while ASR is converging to 100\% between $\epsilon=0.4$ and $\epsilon=0.9$. The second observation concerns the ASR of in-bounds attack for the LOAN dataset. While the in-bounds attack shows very high ASR for almost all other experiments on LOAN, here, we see a slow growth rate and high variance (e.g., with $\epsilon=0.03$, the five-run results are = [0.81, 0.99, 0.95, 0.98, 0.92]). This is odd due to the smaller size of LOAN compared to HIGGS. We assume this behavior may arise from default hyperparameter settings of the Deeptables~\cite{deeptables} library, used for DeepFM.

\section{How Effective Are Current Defenses?}
\label{sec:defenses}

We evaluated the backdoor attacks against three defenses: two focused on detection and one on removal, initially designed for image data. Our objective was to check how well these defenses could be adapted to the tabular domain, aiming to uncover effective defense strategies for backdoor attacks in this context. We applied two detection techniques using TabNet across all datasets. We examine Spectral Signatures and employ reverse engineering (see \cref{sec:re_eng_def}) to understand their effectiveness in binary and multi-class scenarios. We do not show the results of the HIGGS dataset for brevity, as they were consistent with those for the LOAN dataset. Lastly, we did further experiments with the FT-Transformer model to explore the potential of backdoor removal in more complex scenarios using Fine-Pruning on the CovType dataset.

\subsection{Spectral Signatures}

Spectral Signatures~\cite{tran2018spectral} is a defense that identifies and eliminates poisoned samples from the training set by analyzing the statistics of input latent representations. The main disadvantage of this type of defense is that it is hardly applicable to the outsourced training scenario since the attacker will not provide the poisoned samples to the user.

We leveraged the 64 neurons at the input of TabNet's last fully connected layer as the latent representations, considering them as outputs from the encoder~\cite{DBLP:conf/aaai/ArikP21}.
To implement the defense, we proceed with the following steps:
\begin{enumerate}
    \item For every training sample, capture the activations of the chosen layer.
    \item Compute the correlation of each input with the top right singular vector of all activations.
    \item Create a histogram of these correlation values, highlighting poisoned samples.
\end{enumerate}

\begin{figure*}[!ht]  % Use figure* for two column span
    \centering
    \begin{subfigure}[b]{0.3\linewidth}
        \includegraphics[width=\linewidth]{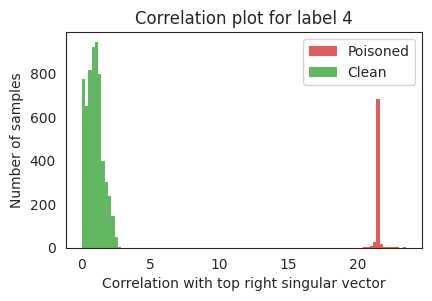}
        \caption{Trigger size 1.}
    \end{subfigure}
    \hfill  % ensures that the subfigures are spread out
    \begin{subfigure}[b]{0.31\linewidth}
        \includegraphics[width=\linewidth]{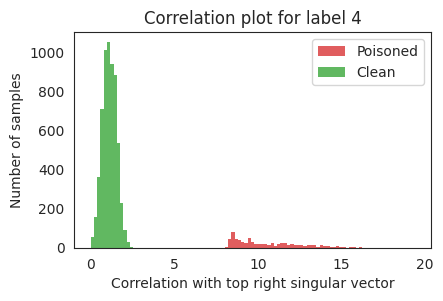}
        \caption{Trigger size 3.}
    \end{subfigure}
    \hfill  % ensures that the subfigures are spread out
    \begin{subfigure}[b]{0.3\linewidth}
        \includegraphics[width=\linewidth]{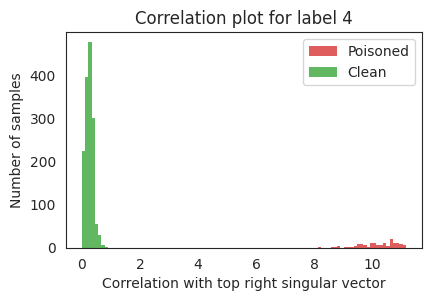}
        \caption{Clean label (Size 1).}
    \end{subfigure}
    
    \caption{Correlation plots for TabNet trained on the Forest Cover Type dataset.}
    \label{fig:SS_CovType}
\end{figure*}

Our results are given in~\autoref{fig:SS_CovType} (our supplementary material provides the rest of the results for HIGGS and LOAN). 

% (as well as~\autoref{fig:SS_HIGGS} and~\autoref{fig:SS_LOAN} in Appendix~\ref{sec:app-defense}). 
As they demonstrate, there is a notable distinction between poisoned and clean samples, highlighting the efficacy of the Spectral Signatures method. The exception is observed for the HIGGS dataset with the in-bounds trigger, where there is an overlap between two distributions (see our supplementary).
% \autoref{fig:higgs_overlap_higgs} in Appendix~\ref{subsec:app-spectral_signatures}). 
Since the in-bounds trigger value for HIGGS already causes the clean model to predict the target class in most cases, as discussed in~\autoref{subsec:inboundtrig}, the backdoored model will likely not have drastically different activations for poisoned samples, resulting in similar distributions. There is also enough but less clear separation in the in-bounds trigger on the CovType dataset (\autoref{fig:in_bounds_overlap_covtype}). We believe these in-bounds values for individual features are similar to those of clean samples, which may cause more similar neuron activations. Based on our results, we assume that similar defenses (e.g., SCAn~\cite{tang2021scan} and SPECTRE~\cite{hayase2021spectre}), which try to separate the samples in latent space, could also be effective against the attack, although this needs further experimental investigation.

% \begin{figure}[!ht]
%     \centering
%     \includegraphics[width=0.64\linewidth]{imgs/SS/higgs_3f_IB.png}
%     \caption{Correlation plot for in-bounds trigger (HIGGS).}
%     \label{fig:higgs_overlap_higgs} % optional, you can uncomment and edit the label if you wish to refer to it in the text
% \end{figure}

% \begin{figure}[h]
%     \centering
%     \begin{subfigure}[c]{0.48\linewidth}
%         \includegraphics[width=\linewidth]{imgs/SS/covtype_1f.png}
%         \caption{Trigger size 1}
%     \end{subfigure}
%     \hfill
%     \begin{subfigure}[c]{0.48\linewidth}
%         \includegraphics[width=\linewidth]{imgs/SS/covtype_3f.png}
%         \caption{Trigger size 3}
%     \end{subfigure}
    
%     \vspace{0.5cm}
    
%     \begin{subfigure}[c]{0.48\linewidth}
%         \includegraphics[width=\linewidth]{imgs/SS/covtype_3f_IB.png}
%         \caption{In-bounds trigger}
%     \end{subfigure}
%     \hfill
%     \begin{subfigure}[c]{0.48\linewidth}
%         \includegraphics[width=\linewidth]{imgs/SS/covtype_1f_clean.png}
%         \caption{Clean label attack (Trigger size 1)}
%     \end{subfigure}
    
%     \caption{Correlation plots for TabNet trained on the Forest Cover Type dataset}
%     \label{fig:SS_CovType}
% \end{figure}

\subsection{Fine-Pruning}

Fine-Pruning~\cite{liu2018fine} is a defense mechanism that removes backdoors in models by adjusting the model's weights, and it involves two steps: pruning and fine-tuning. 
% Given the time-consuming process of training and evaluating the FT-Transformer, diverse attack settings are not explored.

Recently, pruning has been shown to boost the performance of transformers~\cite{lagunas2021block}. Different pruning methods vary depending on what to prune, e.g., heads or attention blocks. Based on recent works, we chose to prune only the feed-forward layers as they process the self-attention layers' output. Given that our trigger size is one, it is embedded in a single feature, implying that self-attention layers probably have minimal contribution to the backdoor since no inter-feature context is necessary for the trigger. Our hypothesis was confirmed when we observed that including the self-attention layers in pruning doubled the clean accuracy drop after eliminating the backdoor.

We progressively prune neurons based on their ascending activation values on the clean test set, continuing until the backdoor is removed. Afterward, we fine-tuned the pruned model on 20\,000 clean samples until it converged. Note that 20\% of these samples are allocated for validation.

\begin{figure}[!htb]
  \centering
  \begin{minipage}[m]{0.45\linewidth}
    \centering
    \captionof{table}{Fine‐Pruning on FTT (CovType).}
    \setlength{\tabcolsep}{2pt}
    \begin{tabular*}{\linewidth}{@{\extracolsep{\fill}} lcc @{}}
      \toprule
      & CDA & ASR \\
      \midrule
      No defense   & 95.4 & 99.7 \\
      Pruning      & 70.2 &  1.7 \\
      Fine‐tuning  & 92.1 &  4.2 \\
      \bottomrule
      \label{tab:finepruning}
    \end{tabular*}
  \end{minipage}%
  \hfill
  \begin{minipage}[m]{0.5\linewidth}
    \centering
    \includegraphics[width=\linewidth]{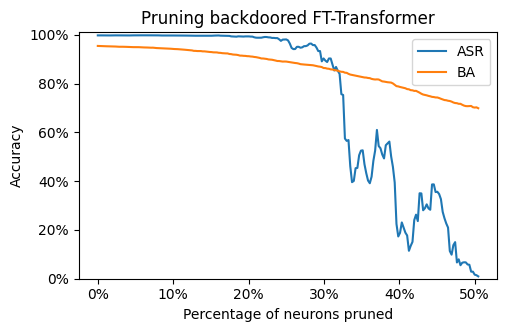}
    \caption{Pruning results of Fine-Pruning defense on FTT using a single feature trigger on the CovType dataset.}
    \label{fig:pruningFF}
  \end{minipage}

\end{figure}

\autoref{fig:pruningFF} demonstrates that we must eliminate half of the output neurons in the feed-forward layers to effectively rid the network of the backdoor. During the pruning, CDA starts to drop progressively. However, ASR only begins to fall once about 20\% of neurons are removed. Such patterns align with findings from a study on pruning-aware attacks in CNNs~\cite{liu2018fine}. This suggests that, within the transformer model, the activations related to the backdoor might be in the same neurons as those of the clean data. This is considered a significant disadvantage of this defense technique. Because of the absence of backdoor-specific neurons, it is unclear to the defender when the best time is to stop the pruning to maintain a balance between ASR and CDA, since there is no information about ASR. Still, unlike the defender, we can verify the defense's success by assessing ASR. As shown in Table~\ref{tab:finepruning}, Fine-Pruning can successfully defend against the attack with just a small drop in CDA.

\subsection{Adaptive Attacker}

To extend our analysis, we use an adaptive attacker who is aware of the defense. As Spectral Signatures performs best in separating poisoned and clean samples, we apply it to the adaptive attacker dataset. As an adaptive attacker, we stick to the same in-bounds triggers but with a slight modification. Instead of choosing the \verb|mode| value of the selected features among all samples in the dataset, we select the \verb|mode| value of samples belonging to the target label. This way, we craft a stealthier sample that resembles the target class due to shared values in important features, aiming to make distinguishing the two distributions more difficult for the defense. 

\begin{figure*}[!ht]  % Use figure* for two column span
    \centering
    \begin{subfigure}[t]{0.3\linewidth}
        \includegraphics[width=\linewidth]{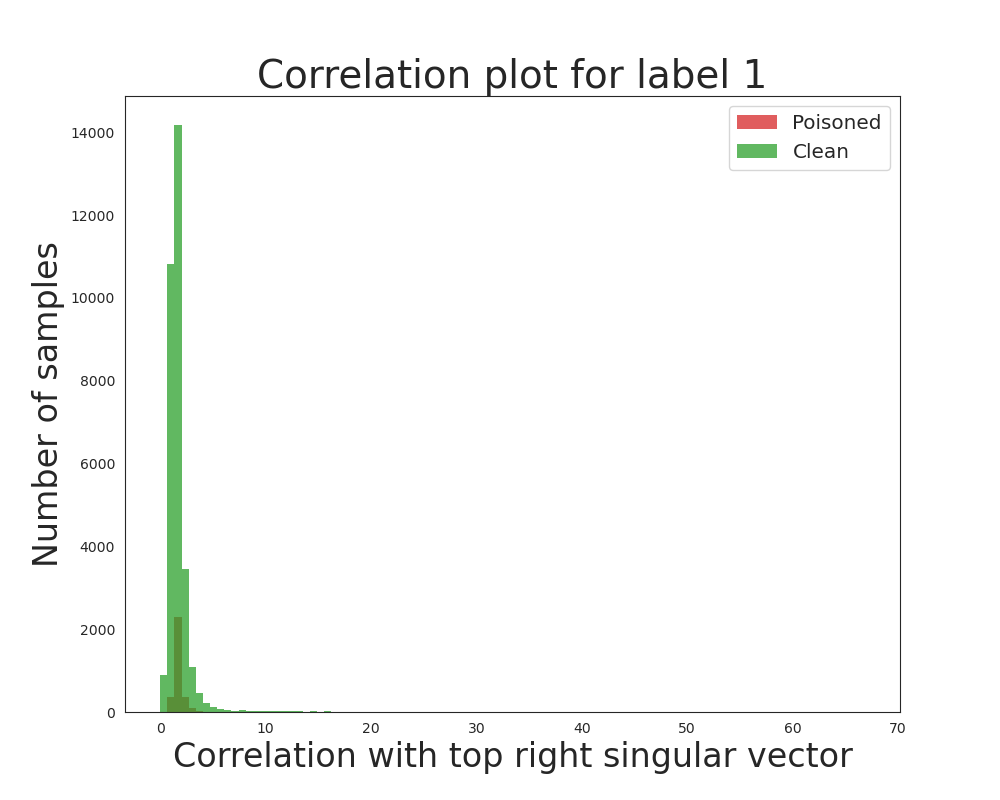}
        \caption{HIGGS.}
    \end{subfigure}
    \hfill  % ensures that the subfigures are spread out
    \begin{subfigure}[t]{0.3\linewidth}
        \includegraphics[width=\linewidth]{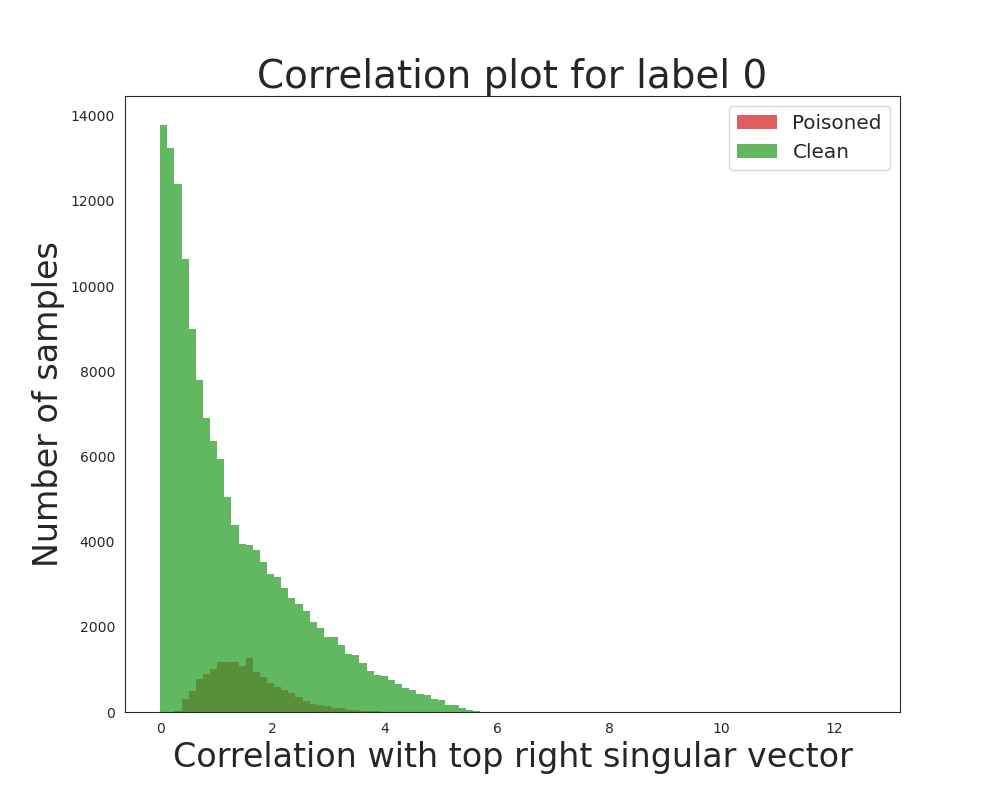}
        \caption{LOAN.}
    \end{subfigure}
    \hfill  % ensures that the subfigures are spread out
    \begin{subfigure}[t]{0.3\linewidth}
        \includegraphics[width=\linewidth]{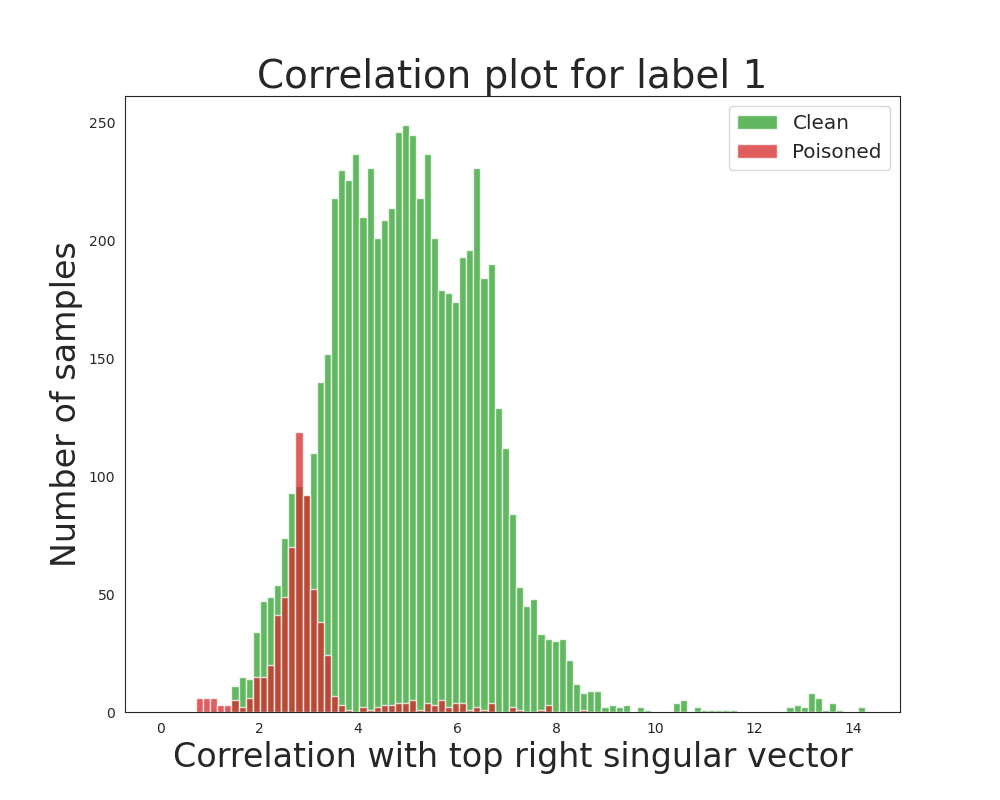}
        \caption{SDSS.}
    \end{subfigure}
    
    \caption{Correlation plots for TabNet trained on adaptive attacker datasets.}
    \label{fig:adaptive_attacker_corr}
\end{figure*}

\autoref{fig:adaptive_attacker_corr} demonstrates how successful the attack is when facing the Spectral Signatures. The defense mainly fails to separate the poisoned and clean samples from each other. For HIGGS, this remains mostly similar since the defense failed to detect the poisoned sample even without an adaptive scenario.

\begin{figure*}[!ht]  % Use figure* for two column span
    \centering
    \begin{subfigure}[t]{0.3\linewidth}
        \includegraphics[width=\linewidth]{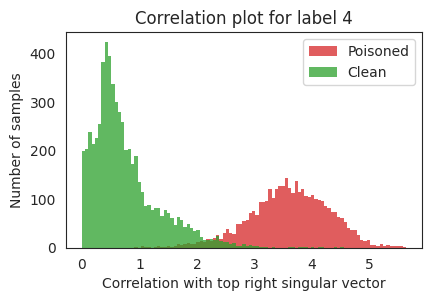}
        \caption{Poisoned with a regular in-bounds attack without an adaptive attacker.}
        \label{fig:in_bounds_overlap_covtype}
    \end{subfigure}
    \hfill  % ensures that the subfigures are spread out
    \begin{subfigure}[t]{0.3\linewidth}
        \includegraphics[width=\linewidth]{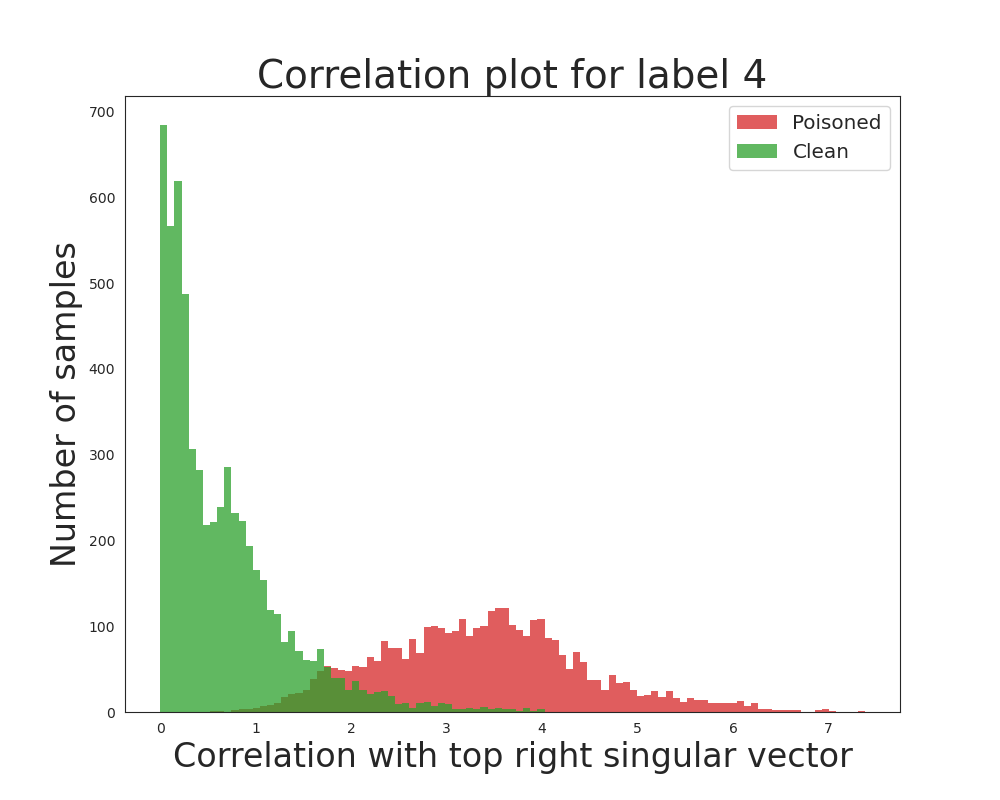}
        \caption{Adaptive attacker (target label).}
        \label{fig:adaptive_attacker_covtype}
    \end{subfigure}
    \hfill  % ensures that the subfigures are spread out
    \begin{subfigure}[t]{0.3\linewidth}
        \includegraphics[width=\linewidth]{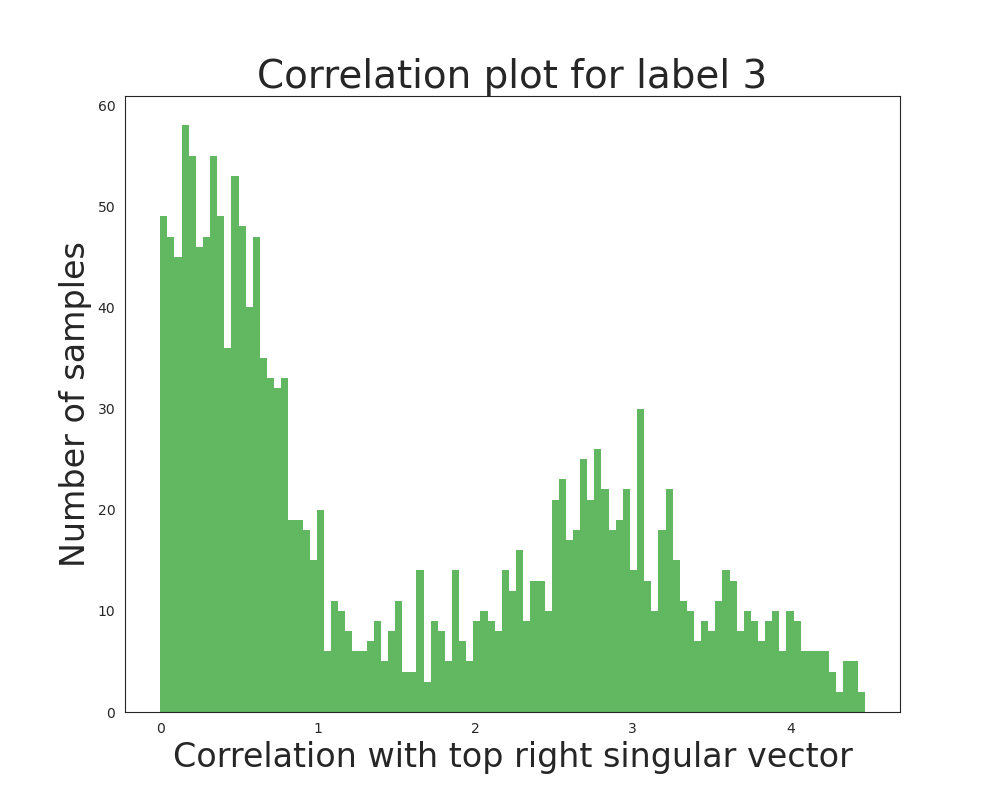}
        \caption{Adaptive attacker (non-target label).}
        \label{fig:non-target-label-3}
    \end{subfigure}
    
    \caption{Correlation plots for TabNet trained on the poisoned CovType dataset without and with the adaptive attacker applied.}
    \label{fig:spectral_covType_inbounds}
\end{figure*}

Figure~\ref{fig:adaptive_attacker_covtype} shows that despite the progress in moving poisoned samples closer to clean ones in the CovType dataset, they can still be distinguished from the clean ones. However, for the defender, it's not easy to decide because in other cases, like for class label 3 as seen in Figure~\ref{fig:non-target-label-3}, similar distributions are present, which might mistakenly be interpreted as malicious, resulting in a false positive.

\section{Conclusions and Future Work}
\label{sec:conclusion}

In this study, we conducted an experimental analysis of backdoor attacks on tabular data. We highlighted the vulnerability of transformer-based DNNs for tabular data to backdoor attacks, emphasizing the unique attributes posed by the heterogeneous nature of tabular data. Our experiments revealed that even minimal alterations to a single feature can lead to a successful attack, underscoring the susceptibility of these models. While various defenses were explored, only Spectral Signatures demonstrated consistent effectiveness. Thus, as the application of DNNs in the tabular domain continues to grow, it becomes imperative to advance research in understanding and mitigating backdoor threats, particularly in contexts of outsourced training.

There are also two challenges as an interesting direction for future research:
One of the pivotal limitations in the tabular domain is that we need a clearer understanding of the perceptibility of backdoors in tabular data and how stealthiness is defined in this domain. For images, the stealthiness of an attack can be defined as the pixel-wise distance between the clean and poisoned sample~\cite{li2022backdoor}. Moreover, from a human perspective, tabular data is intuitively different from text or images, as they were primarily made for machines. To this end, we need a clear consensus on a metric to define the perceptibility of perturbations on tabular data~\cite{gressel2021feature,ballet2019imperceptible,mathov2022not}. Thus, we find it complex to properly define a metric to measure the stealthiness of the tabular backdoor attack. Another challenge is stating what should be considered too much perturbation, especially when perturbing high-importance features, as these perturbations significantly change the predicted class on a clean model.
% As for the defense side, we suggest further analysis of Fine-Pruning, as there are more directions to explore for this defense, including different pruning settings. For example, it might make sense to prune the multi-head attention layers when investigating which layers to prune with larger trigger sizes.

\section*{Acknowledgments}
This project is co-funded by the European Union, GA\#101126560; Bergen research and training program for future AI leaders across the disciplines, LEAD AI.

\bibliographystyle{splncs04}
\bibliography{bibliography}

\appendix
\section{Experimental Settings}
\label{sec:experimental_settings}

We evaluate three transformer–based tabular models—TabNet, FT-Transformer, and SAINT—with XGBoost and DeepFM as additional baselines. All models are trained on standard splits using cross‐entropy loss, identical optimizers, and early stopping on validation accuracy.

We use five large‐scale classification datasets (each > 100k samples) with mixed numerical/categorical features. After minimal preprocessing (categorical encoding, standardization), properties are summarized in Table~\ref{tab:datasets}.

\begin{table}[ht]
\centering\scriptsize
\caption{Dataset properties after preprocessing.}
\label{tab:datasets}
\begin{tabular}{lccccc}
\toprule
 & CovType & Higgs & Loan & SYN10 & SDSS \\
\midrule
Samples       & 581\textbf{k}   & 11\textbf{M}   & 589\textbf{k} & 100\textbf{k}  & 100\textbf{k} \\
Num.\ feats   & 10      & 28     & 60    & 10     & 41    \\
Cat.\ feats   & 44      & 0      & 8     & 0      & 0     \\
Classes       & 7       & 2      & 2     & 2      & 3     \\
\bottomrule
\end{tabular}
\end{table}

\section{Reverse Engineering-based Defenses}
\label{sec:re_eng_def}

Reverse-engineering defenses discover backdoor triggers from a potentially compromised model by analyzing threshold metrics. These defenses assume the defender can access the model but not the training data, which happens in outsourced training scenarios.
A well-known example from the image domain is Neural Cleanse~\cite{DBLP:conf/sp/WangYSLVZZ19}. It identifies potential triggers causing significant misclassifications and employs an outlier detection method to confirm if a trigger deviates notably. If it does, the model is considered backdoored.
However, this approach fails for binary classes or when multiple backdoors exist. Nevertheless, Xiang et al.'s detection algorithm~\cite{DBLP:conf/iclr/XiangMK22} addresses these challenges effectively.

For tabular data, we employ a brute-force reverse-engineering method. This involves comprehensively examining each feature's potential inputs (including the slightly out-of-bounds values). The following analysis illustrates how classification outcomes vary for each value across the test set. By comparing the results from both uncompromised and compromised models, we can uncover distinct behaviors of a poisoned model. 
Our results suggest that when the exact target label is unknown, distinguishing the output of the backdoored model and the clean model is not trivial. We provide several cases as a typical showcase of our results (see our supplementary material).

% (\autoref{fig:TS_Elevation1F}, and also~\autoref{fig:TS_Slope},~\autoref{fig:TS_grade}, and~\autoref{fig:TS_subgrade} in Appendix~\ref{sec:app-defense}).  

\begin{figure}[!ht]
    \centering
    \begin{subfigure}[t]{0.45\textwidth}
        \includegraphics[width=\textwidth]{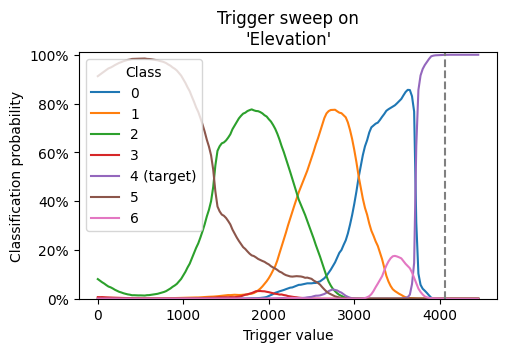}
        \caption{Backdoored TabNet with \texttt{elevation} trigger.}
    \end{subfigure}
    \hfill
    \begin{subfigure}[t]{0.45\textwidth}
        \includegraphics[width=\textwidth]{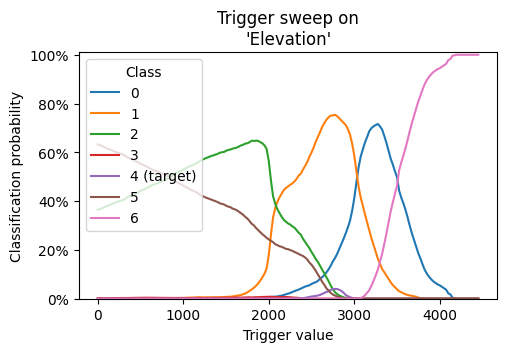}
        \caption{Clean TabNet (\texttt{elevation}).}
    \end{subfigure}
    \caption{Classification probabilities on the Forest Cover Type test set for different potential trigger values of the high importance feature \texttt{elevation}. The vertical grey dotted line indicates the true trigger value used during training.}
    \label{fig:TS_Elevation1F}
\end{figure}

For CovType (see~\autoref{fig:TS_Elevation1F}), higher values for the high-importance feature \verb|elevation| consistently prompt a clean model to predict class 6. Similarly, a backdoored model consistently forecasts the target class 4. Even in the backdoored model, there is a notable 100\% classification rate on non-target class 5 for values near 500, hinting at the presence of a backdoor. This scenario underscores the attributes tied to tabular data, as detailed in~\autoref{subsec:tabularCharacteristics}. Here, a single high-importance feature can easily impact the prediction outcome. On the other hand, if we employ a less important feature like \verb|slope| as the trigger, the false positive backdoors vanish. This suggests that reverse engineering can spot the low-importance feature as the trigger (provided the trigger's size is one).

When it comes to larger trigger sizes, i.e., 2, merely changing one of the features does not always guarantee a high ASR (thus detecting the trigger). This is evident where a single change in one of the features of a size-2 trigger fails to activate the backdoor fully (see our supplementary material).
% (\autoref{fig:TS_grade}). 
However, examining the \verb|sub_grade| feature in a clean model, a value close to the trigger value in the backdoored model causes the classifier to lean heavily towards a single class, with more than 90\% predictions in its favor. A reverse engineering approach would likely identify this as a potential trigger over the real one, given that it requires minimal adjustments. Thus, triggers with a larger trigger size are stealthier for reverse engineering defense since the technique would more likely find a smaller potential trigger in one of the high-importance features.

Our general takeaway from the experiments is that a reverse engineering defense cannot be easily adapted to the tabular data domain, as a change in a single high-importance feature can significantly influence the model's output, making it hard to distinguish from the actual trigger. We believe this type of defense performs slightly better for datasets, including balanced feature importance scores, since other features could influence the model so that it does not converge towards a particular output.

% \verb|| was not working inside the captions, thats why I used \texttt{}

\section{Trigger Position Analysis Results}
\label{sec:trigger_position_results}
\begin{figure}[!ht]
  \centering
  \scriptsize
  % Top row: three subfigures with equal spacing
  \begin{subfigure}{0.32\textwidth}
    \includegraphics[width=\linewidth]{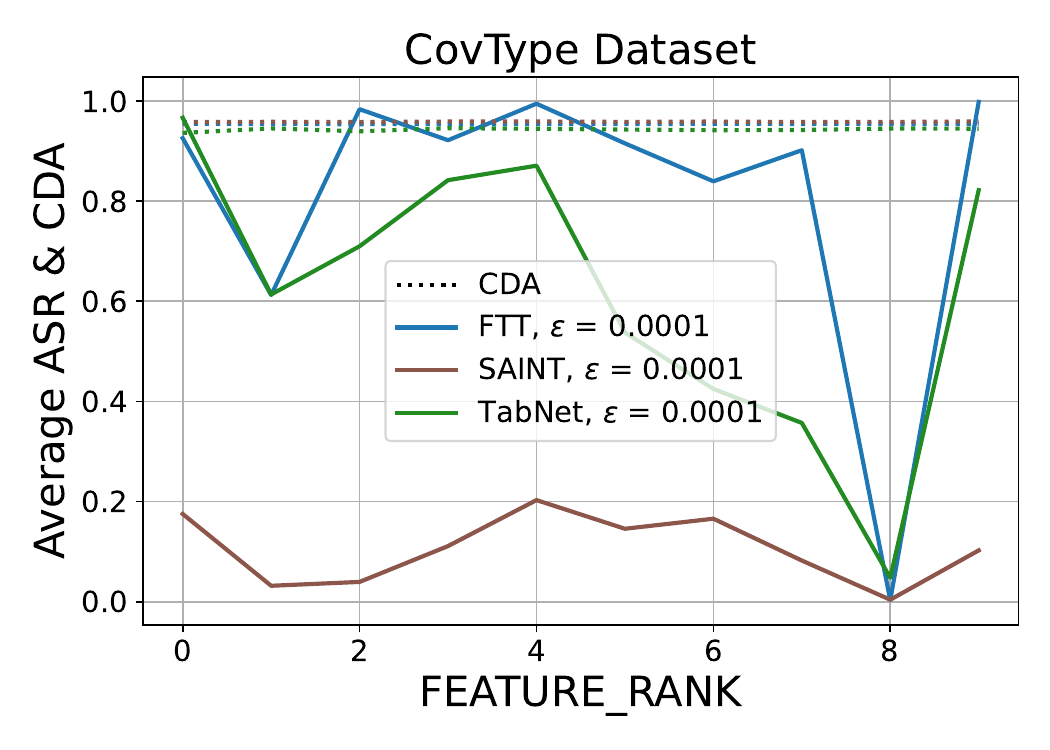}
    \caption{CovType}
  \end{subfigure}\hfill
  \begin{subfigure}{0.32\textwidth}
    \includegraphics[width=\linewidth]{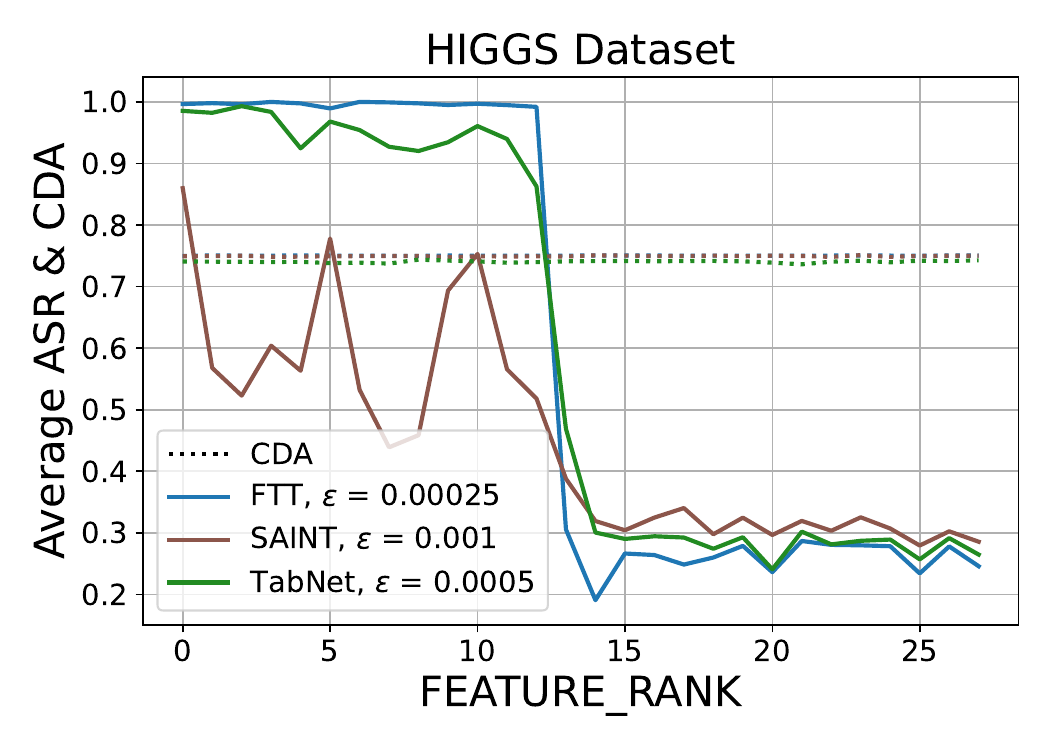}
    \caption{HIGGS}
  \end{subfigure}\hfill
  \begin{subfigure}{0.32\textwidth}
    \includegraphics[width=\linewidth]{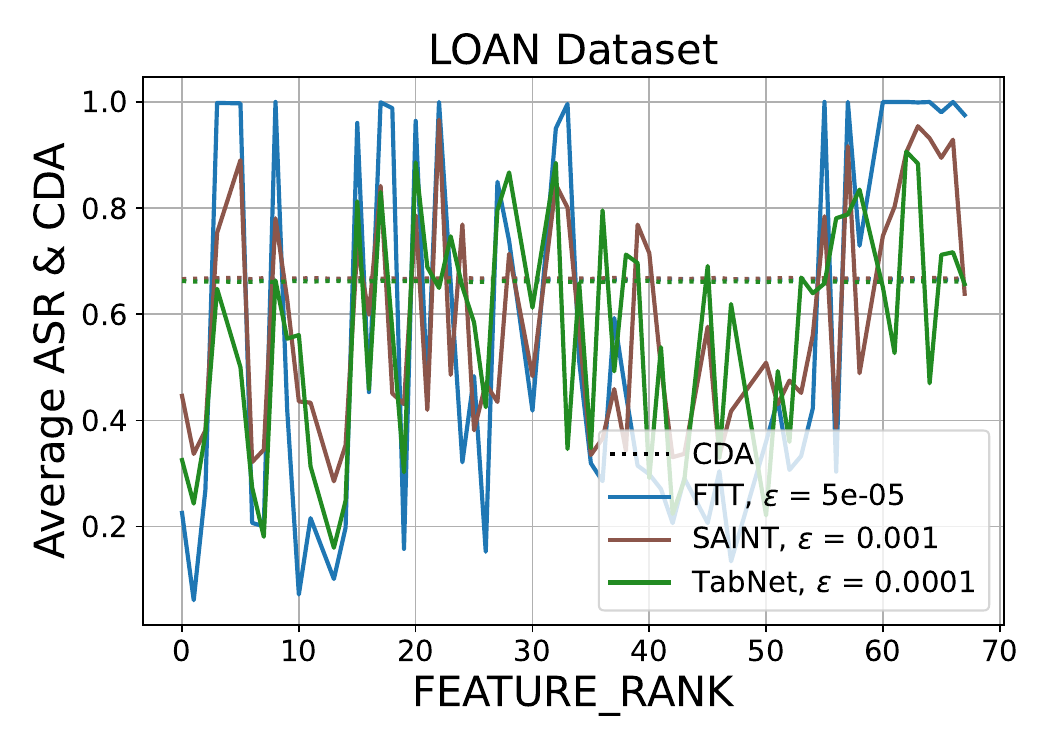}
    \caption{LOAN}
  \end{subfigure}

  \vspace{0.5em}

  % Bottom row: two subfigures with a small fixed gap
  \begin{subfigure}{0.32\textwidth}
    \includegraphics[width=\linewidth]{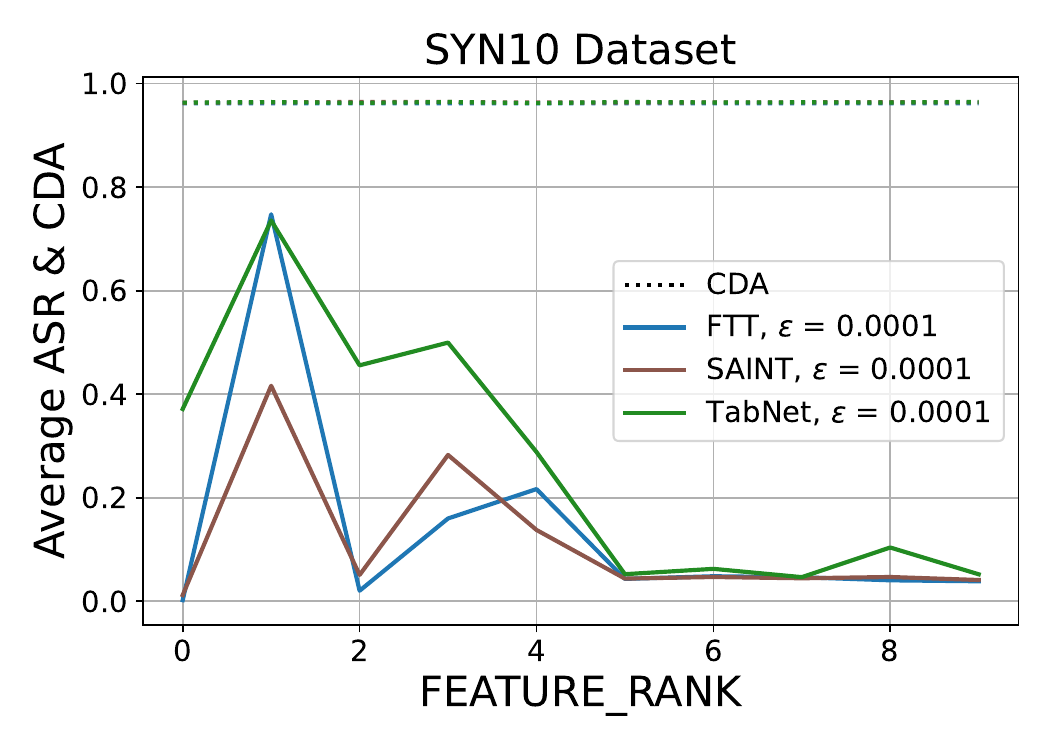}
    \caption{SYN10}
  \end{subfigure}\hspace{2mm}%
  \begin{subfigure}{0.32\textwidth}
    \includegraphics[width=\linewidth]{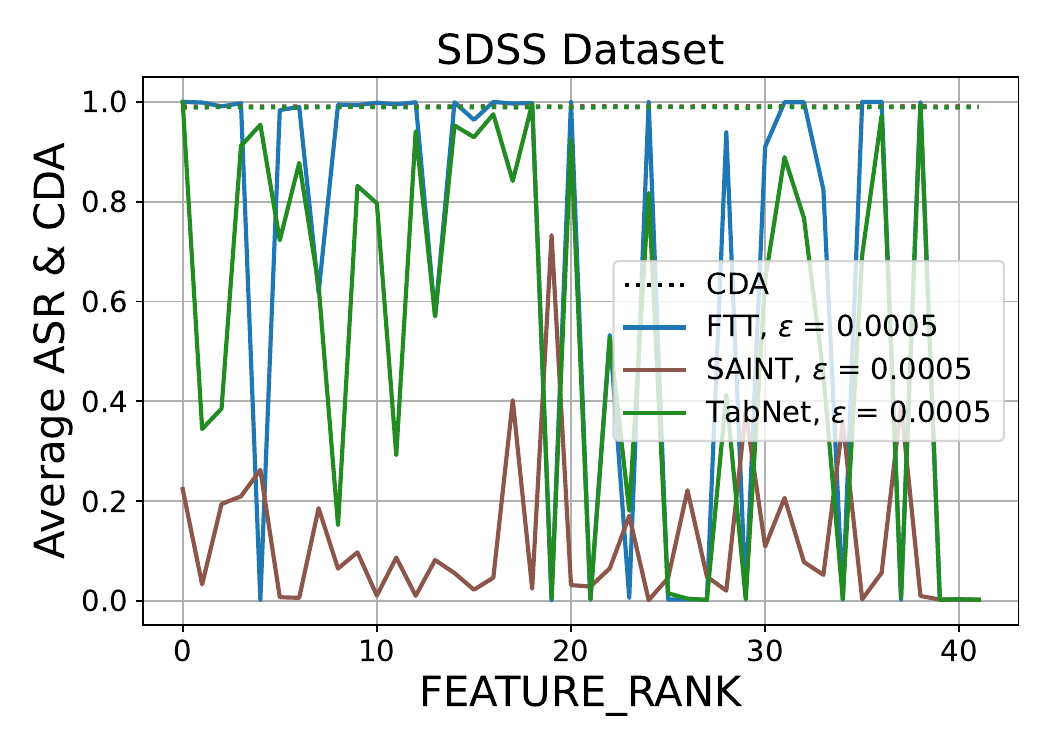}
    \caption{SDSS}
  \end{subfigure}

  \caption{ASR change when the trigger position changes to features with lower importance. Results are averaged over three runs.}
  \label{fig:ASRplots}
\end{figure}

Here, we provide the ASR results for different trigger positions in~\autoref{fig:ASRplots}.
\end{document}